\documentclass[%
 reprint,
superscriptaddress,
 amsmath,amssymb,
prx,
floatfix,
]{revtex4-2}

\usepackage{graphicx}
\usepackage{dcolumn}
\usepackage{bm}
\usepackage{xcolor}

\usepackage{bbold}
\usepackage{xparse}
\usepackage{hyperref}
\usepackage{siunitx}
\usepackage{cleveref}
\usepackage{braket,bigints} 

\usepackage{bbm} 

\renewcommand\bra[1]{{\langle{#1}|}}
\makeatletter
\renewcommand\ket[1]{%
\@ifnextchar\bra{\k@t{#1}\!}{\k@t{#1}}%
}
\newcommand\k@t[1]{{|{#1}\rangle}}
\makeatother

\usepackage{color}
\definecolor{mygreen}{rgb}{0,0.5,0}
\definecolor{mygrey}{rgb}{0.5,0.5,0.5}
\definecolor{myred}{rgb}{0.75,0,0}
\definecolor{myblue}{rgb}{0,0,0.75}
\definecolor{mymagenta}{cmyk}{0,1,0,0.12}
\definecolor{mycyan}{cmyk}{1,0,0,0.12}
\definecolor{myorange}{rgb}{1.,0.5,0}
\definecolor{myviolet}{rgb}{0.6,0.15,0.6}
\definecolor{mybrown}{cmyk}{0,0.50,1,0.41}

\usepackage{ulem}

\newcommand{\supin}{^{(\mathrm{in})}}
\newcommand{\supout}{^{(\mathrm{out})}}
\usepackage{mathtools}

\DeclareMathOperator{\var}{var}
\newcommand{\NAtoms}{N_\mathrm{at}}

\begin{document}

\preprint{APS/123-QED}

\title{Effects of spin-exchange collisions on the fluctuation spectra of hot alkali-metal-metal vapors}

\newcommand{\ICFO}{ICFO - Institut de Ci\`encies Fot\`oniques, The Barcelona Institute of Science and Technology, 08860 Castelldefels (Barcelona), Spain}
\newcommand{\ICREA}{ICREA - Instituci\'{o} Catalana de Recerca i Estudis Avan{\c{c}}ats, 08010 Barcelona, Spain}
\newcommand{\ITE}{Institute of Electronic Structure and Laser, Foundation for Research and Technology, 71110 Heraklion, Greece}
\newcommand{\UOC}{Department of Physics and Institute of Theoretical and Computational Physics, University of Crete, 70013 Heraklion, Greece}
\newcommand{\Hangzhou}{Department of Physics, Hangzhou Dianzi University, Hangzhou 310018, China}

\author{K. Mouloudakis}
\email{kostas.mouloudakis@icfo.eu}
\affiliation{\ICFO}
\author{G. Vasilakis}
\email{gvasilak@iesl.forth.gr}
\affiliation{\ITE}
\author{V. G. Lucivero}
\affiliation{\ICFO}
\author{J. Kong}
\affiliation{\Hangzhou}
\author{I. K. Kominis}
\affiliation{\UOC}
\affiliation{Quantum Biometronics PC, 71409 Heraklion, Greece}
\author{M. W. Mitchell}
\email{morgan.mitchell@icfo.eu}
\affiliation{\ICFO}
\affiliation{\ICREA}




\date{\today}

\begin{abstract}

We present a first-principles analysis of the noise spectra of alkali-metal-metal vapors in and out of the spin-exchange-relaxation-free (SERF) regime, and we predict non-intuitive features with a potential to further improve the sensitivity of SERF media, and which must be taken into account in their use in quantum optical applications. Studying the process of spin-noise spectroscopy (SNS), we derive analytic formulas for the observable noise spectra, and for the correlation functions among different hyperfine components, which give additional insight into the spin dynamics. The analytic results indicate a variety of distortions of the spin-noise spectrum relative to simpler models, including a broad spectral background that mimics optical shot noise, interference of noise contributions from the two ground-state hyperfine levels, noise reduction at the spin-precession frequency, and ``hiding'' of spin-noise power that can introduce a systematic error in noise-based calibrations, e.g., for spin-squeezing experiments.

\end{abstract}

\maketitle


\section{Introduction}
\label{sec:Introduction}

Hot alkali-metal vapors are a workhorse medium for many atomic sensing tasks, including magnetometry \cite{budker2007optical}, magnetic gradiometry \cite{LimesPRAppl2020, LuciveroPRAppl2021}, comagnetometry \cite{PhysRevLett.124.170401,RomalisVas-comag}, alkali-metal--noble-gas gyroscopes \cite{WalkerBook2016,FangSensors2012},  and quantum memories \cite{Shaham2022}, as well as tabletop searches for new  physics \cite{Afach2021,PhysRevLett.125.201802,pustelny2013global}. Recent works have also shown that these media support non-classical correlations \cite{MitchellNatureCom,KuzmichSpinSqueezingQND,VasilakisPolzik2015,PhysRevA.102.012822,PRXQuantum.3.010305}, allowing quantum enhancement techniques, e.g., squeezing of light or of spins, to improve their performance \cite{Troullinou,Julsgaard2001,Sherson2006,TrutleinHe3}.
Detailed models for the behavior of such devices include intra- and inter-atomic spin dynamics, response to probe and pumping light, and external fields \cite{seltzerthesis, SavukovRomalis}. In these models, spin-exchange (SE) collisions play an important role, and in practice they often limit the spin-coherence time and consequently the performance of these devices. 

SE collisions \cite{Grossette1,Grossette2} between pairs of alkali-metal atoms, together with the intra-atomic spin dynamics and coupling to external fields, lead to non-intuitive relationships among atomic number density, magnetic field, and coherence time \cite{PhysRev.153.132}.  At low densities or high fields, SE collisions are a decoherence mechanism, causing spin coherence times to decrease with increasing density. In the so-called spin-exchange relaxation-free (SERF) regime of high density and low field, however, frequent SE collisions decouple the net spin from decoherence via the external field, and  coherence times increase with increasing density \cite{PhysRevLett.31.273, PhysRevA.16.1877}.

Spin-noise spectroscopy (SNS)  provides a window into the non-intuitive spin dynamics of dense atomic vapors \cite{lucivero2017sensitivity}. SNS measures the spin fluctuations of an atomic spin ensemble, driven only by random influences such as collisions and diffusion of spins into and out of the system. These spin dynamics show correlations that reflect the magnetic resonance spectral response \cite{PhysRevA.104.063708,BruunPRL2009, MullerAPL2010}, and they are acquired without excitation of the medium. In particular, the spin-noise spectrum of an unpolarized vapor reveals the vapor's transition from spin-exchange and SERF regimes, including the non-intuitive linewidth behavior \cite{MitchellNatureCom,PhysRevA.104.063708,katsoprinakis}.

Despite recent experimental and theoretical study of the noise properties of SERF-regime vapors, nearly all work has focused on the most prominent feature, which is a Lorentzian-like peak at the Larmor frequency (taking into account  nuclear slowing). In this work, we calculate from first principles the full spin-noise spectrum of a hot alkali-metal vapor, including the effects of hyperfine, Zeeman, and spin-exchange on the vapor, and the effects of probe detuning on the coupling to the different ground hyperfine states $F_a$ and $F_b$. We find features not previously remarked, including a spectrally broad spin-noise background that mimics optical shot noise, departure of the spin-noise spectrum from the Lorentzian shape, a probe-detuning-dependent ``interference'' of noise contributions from $F_a$ and $F_b$, and in some circumstances, a significant reduction of spin-noise at the precession frequency. We also describe how, in typical experimental conditions, the integrated spin-noise power can deviate by as much as 20\% from the predictions of simpler models. This can significantly affect the use of spin-noise of unpolarized vapors to benchmark quantum noise, e.g., in quantum sensing.

The paper is organized as follows.  In Sec. \ref{sec:Model}, we present the statistical model for the spin dynamics, first in terms of a master equation describing the evolution of the single-atom density matrix $\rho$, and then in terms of an equivalent stochastic Bloch equation that governs the collective spin observables  $\mathbf{\hat{F}}^{a}$ and  $\mathbf{\hat{F}}^{b}$. The spin readout, by Faraday rotation, is described by input-output relations for the probe beam's Stokes operators, including intrinsic polarization noise. In Sec. \ref{sec:Correlations}, we derive analytic results for correlation functions and spectra. In Sec. \ref{sec:PredictedPhenomena}, we predict several observable SNS phenomena based on the analystic results.

\section{Model}
\label{sec:Model}

\subsection{Spin dynamics -- master equation}
\label{sec:MasterEquation}

In the presence of a constant magnetic field $\mathbf{B}$, the angular momenta in the two hyperfine multiplets precess in opposite directions. Spin-exchange collisions transfer the atoms between the two hyperfine manifolds resulting in a spin-dephasing evolution captured by the single-atom density matrix equation \cite{SavukovRomalis}
\begin{equation}
 \frac{d \rho}{d t} = -\frac{i}{\hbar} [A_{\mathrm{hf}} \mathbf{\hat{I}} \cdot \mathbf{\hat{s}} +g_s \mu_{_B} \mathbf{\hat{s}} \cdot \mathbf{B}, \rho] + R_{\mathrm{se}} [\hat{\varphi} (\mathbbm{1}+4 \langle \mathbf{\hat{s}} \rangle \cdot \mathbf{\hat{s}})- \rho], \label{eq:density-matrix}
 \end{equation} 
where $A_{\mathrm{hf}}$ is the hyperfine coupling constant, $g_s \approx 2$ is the electron g-factor, $\mu_{_B} \approx 9.27 \times 10^{-24}$ \si{\joule\per\tesla} is the Bohr magneton, $R_{\mathrm{se}}$ is the spin-exchange rate, $\langle \mathbf{\hat{s}} \rangle= \mathrm{Tr}[\rho \mathbf{\hat{s}}]$ is the expectation value of the electronic spin in the state $\rho$ and  $\hat{\varphi} = \sum_{i=0}^3 \hat{s}_i \rho \hat{s}_i$ is the electron-spin-depolarized density operator. Here, $\hat{s}_0 \equiv \mathbbm{1}/2$ and $\{\hat{s}_1,\hat{s}_2,\hat{s}_3\} = \{\hat{s}_x,\hat{s}_y,\hat{s}_z\}$. The steady-state solution of Eq.\eqref{eq:density-matrix} is the thermal spin state $\rho_{\mathrm{eq}} = e^{\bm{\beta} \cdot \mathbf{\hat{f}}}/\mathrm{Tr}[e^{\bm{\beta} \cdot \mathbf{\hat{f}}}]$ where $\mathbf{\hat{f}} \equiv \mathbf{\hat{f}}^a + \mathbf{\hat{f}}^b$ is the total angular momentum in the ground-state, with $a=I+1/2$ and $b=I-1/2$, and $\bm{\beta}$ is the inverse spin-temperature related to the spin-polarization $\mathbf{P} \equiv \langle \mathbf{\hat{s}} \rangle/s$ as $\bm{\beta}~||~\mathbf{P}$, $\beta_j= \mathrm{ln}[ (1+P_j)/(1-P_j)]$ with $j \in \{ x,y,z\}$ \cite{Appelt,HapperRev}. In Eq.\eqref{eq:density-matrix} we have ignored the SE frequency shift,  which is negligible for a low-polarized vapor and the small interaction of the nuclear spin with the magnetic field. Finally, regarding the angular momentum degrees of freedom, the dimension of the Hilbert space in the ground-state and subsequently of the operators entering Eq.\eqref{eq:density-matrix} is $d\equiv (2s+1)(2I+1)$.

The applied magnetic field exerts a torque both on the electron and the nuclear spin. Since the nuclear g-factor is significantly smaller than electron's $g_s$, the angular momenta in the two hyperfine levels precess according to $d\langle \mathbf{\hat{f}^{\alpha}}\rangle/dt\approx \langle \mathbf{\hat{s}^{\alpha}}\rangle \times \gamma_0 \mathbf{B}$ with $\alpha \in\{ a,b\}$. The mean value of the electron spin operator, projected in the two hyperfine levels, is related to the corresponding total angular momentum operator through the relations: $\langle \mathbf{\hat{s}^{a}}\rangle =  \langle \mathbf{\hat{f}^{a}}\rangle/(2I+1)$ and $\langle \mathbf{\hat{s}^{b}}\rangle =-  \langle \mathbf{\hat{f}^{b}}\rangle/(2I+1)$, respectively \cite{optical_mag}.

Eq.~\eqref{eq:density-matrix} is a non-linear ordinary differential equation describing the evolution of the single-atom density matrix. The term containing $\langle \hat{\mathbf{s}}(t) \rangle$ makes the equation non-linear, and its general solution is difficult without resort to numerical techniques.  In SNS scenarios with unpolarized ensembles, however, the vapor is at most weakly-polarized, i.e., $|\langle \hat{\mathbf{s}} \rangle| \ll 1/2$. This allows the  master equation to be linearized, as in  \cite{PhysRevA.16.1877}, to obtain analytic solutions. More details can be found in Appendix \ref{sec:spin-dyn}. In what follows, we consider only this weakly-polarized regime.

\subsection{Spin dynamics --  Bloch equations}
\label{sec:BlochEquations}

Eq.~\eqref{eq:density-matrix}  describes well the evolution of the mean spin polarization, but does not give information about fluctuations about this mean. To study these, it is convenient to introduce collective spin observables   $\mathbf{\hat{F}}^{\alpha}\equiv\sum_{i=1}^{\NAtoms } \mathbf{\hat{f}}_{i}^{\alpha}$, $\alpha \in \{ a,b \}$, where $\hat{\mathbf{f}}_i^\alpha$ is the single-atom spin observable of the $i$th atom, and $\NAtoms$ is the number of atoms. Observable signals and their moments can then be computed as, e.g.,  $\langle \mathbf{\hat{F}}^{\alpha} \rangle =\mathrm{Tr}[\tilde{\rho} \mathbf{\hat{F}}^{\alpha}]$ or  $\langle \hat{F}_i^{\alpha}{\hat{F}}_j^{\beta} \rangle =\mathrm{Tr}[\tilde{\rho} {\hat{F}}_i^{\alpha}{\hat{F}}_j^{\beta}]$, where $\tilde\rho$ is the many-atom state.

As described in \cite{PhysRevA.16.1877,PhysRevA.103.043116,katz2015coherent,PhysRevLett.110.263004,ofer} and in Appendix \ref{sec:spin-dyn}, in the low-polarization regime Eq.~\eqref{eq:density-matrix} implies the following differential equations for $\langle \mathbf{\hat{F}^a} \rangle$ and $\langle \mathbf{\hat{F}^b} \rangle$, known as the hyperfine Bloch equations
\begin{eqnarray}
\frac{d \langle \mathbf{\hat{F}^a} \rangle}{dt} &=& \langle \mathbf{\hat{F}^a} \rangle \times \gamma_0 \mathbf{B} -R_{\mathrm{se}} \kappa_{aa} \langle \mathbf{\hat{F}^a} \rangle +R_{\mathrm{se}}  \kappa_{ab} \langle \mathbf{\hat{F}^b} \rangle,
\label{eq:DiffEqA}\\
\frac{d \langle \mathbf{\hat{F}^b} \rangle}{dt}&=&- \langle \mathbf{\hat{F}^b} \rangle \times \gamma_0 \mathbf{B} -R_{\mathrm{se}}  \kappa_{bb} \langle \mathbf{\hat{F}^b} \rangle +R_{\mathrm{se}}  \kappa_{ba} \langle \mathbf{\hat{F}^a} \rangle.
\hspace{6mm}
\label{eq:DiffEqB}
\end{eqnarray}

Here $\gamma_0=\gamma_e/(2I+1)$ is the atomic gyromagnetic ratio with $\gamma_e = g_s \mu_{_B}/ \hbar \approx 2 \pi \times 2.8 \times 10^4$ \si{\mega\hertz\per\tesla} being the electron gyromagnetic ratio, when neglecting the small contribution of the nuclear Zeeman term, $\omega_0=\gamma_0 B$ the linear Larmor frequency resulting from the diagonalization of the Breit-Rabi Hamiltonian $\hat{H}=A_{\mathrm{hf}} \mathbf{\hat{I}} \cdot \mathbf{\hat{s}} +g_s \mu_{_B} \mathbf{\hat{s}} \cdot \mathbf{B}$ at low magnetic fields and $R_{\mathrm{se}}= n \sigma_{\mathrm{se}} u$ is the spin-exchange rate with $\sigma_{\mathrm{se}} \approx 2 \times 10^{-14}$ cm$^2$ being the spin-exchange cross-section and $u$ the mean relative velocity of the pairwise collisions. The coefficients $\kappa_{\alpha\beta}$ with $\alpha,\beta \in \{a,b\}$ describe the relaxation and the coupling between the two hyperfine multiplets. Following Appendix \ref{sec:spin-dyn}, for $a=I+1/2$ and $b=I-1/2$ it is useful to express the coefficients $\kappa_{\alpha \beta}$ in terms of the nuclear spin
\begin{eqnarray}
\kappa_{ba}  &=&\kappa_{aa}  = \frac{2}{3}\frac{I(2I-1)}{(2I+1)^2},
\\
\kappa_{ab}&=& \kappa_{bb}= \frac{2}{3}\frac{(2I+3)(I+1)}{(2I+1)^2}.
\end{eqnarray}
From Eqs.~\eqref{eq:DiffEqA} and \eqref{eq:DiffEqB} we obtain that $d(\langle \mathbf{\hat{F}^a} \rangle+\langle \mathbf{\hat{F}^b} \rangle)/dt$, does not depend on the $R_\mathrm{se}$.  Moreover, at $\mathbf{B}=0$ the total angular momentum of the two ground-state levels $\langle \mathbf{\hat{F}^a} \rangle+\langle \mathbf{\hat{F}^b} \rangle$ is a constant of motion.

The coupled system of differential equations in Eqs.~\eqref{eq:DiffEqA} and \eqref{eq:DiffEqB} captures the peculiar features of the spin-exchange interaction in a DC magnetic field, specifically, the narrowing of the magnetic resonance spectrum in the SERF regime, as well as, the frequency shift of the resonance frequency $\omega_0$ towards lower frequencies. Both these effects are a consequence of the rapid transfer of angular momentum between the two hyperfine multiplets and start to become significant when the spin-exchange rate is comparable or exceeds the spin-precession frequency $\omega_0$.

\subsection{Spin dynamics --  stochastic Bloch equations}
\label{sec:StochasticBlochEquations}

Eqs.~\eqref{eq:DiffEqA} and \eqref{eq:DiffEqB}, which describe the evolution of the statistical averages $\langle \mathbf{\hat{F}}^\alpha \rangle$, predict a relaxation toward a zero mean polarization. This mean does not, however,  capture the full dynamics of $\mathbf{\hat{F}}^\alpha$, which fluctuate about their averages in a correlated fashion. To account for spin-noise fluctuations in the thermal spin state we first suppose that the magnetic field is transverse to the probe beam, namely $\mathbf{B}=(0,B,0)^T$ along the $\mathbf{\hat{y}}$-direction. Additionally, we define the angular momentum vector $\mathbf{\hat{X}}(t)\equiv[\hat{F}^{a}_x(t),\hat{F}^{a}_y(t),\hat{F}^{a}_z(t),\hat{F}^{b}_x(t),\hat{F}^{b}_y(t),\hat{F}^{b}_z(t)]^T$ and the corresponding noise vector 
$d {\cal{\mathbf{\hat{W}}}}(t)\equiv[d\mathcal{\hat{W}}^{a}_x(t),\ldots,d\mathcal{\hat{W}}^{b}_z(t)]^T$.
Here, $d{\mathcal{\hat{W}}}^{\alpha}_{j} (t)$ with $j \in \{x,y,z\}$ and $\alpha \in \{a,b\}$ represents white Gaussian noise with zero mean and variance $dt$. In addition, we assume that the angular momentum vector $\mathbf{\hat{X}} (t)$ satisfies the stochastic differential equation $d \mathbf{\hat{X}} (t) = A \mathbf{\hat{X}}(t) dt+ Q  d{\cal{\mathbf{\hat{W}}}}(t)$, namely, we assume that the angular momentum operators are stochastic variables and we augment the deterministic differential equations with Gaussian source noise terms associated with the spontaneous fluctuations of the collective angular momenta in the two hyperfine multiplets \cite{PhysRevLett.120.040503} \footnote{An alternative approach, considering the dynamical matrix $A$ to be stochastic, and thus describing the spin dynamics with a random ordinary differential equation (RODE), appears to be equivalent to this method \cite{HanBook2017}}. The additive Gaussian noise terms entering the equation above are a consequence of the central limit theorem since the collective spin variable is randomly affected by many independent processes, e.g., many pairwise collisions with different impact parameters \cite{PRXQuantum.3.010305}.  

$Q$ is the noise-strength matrix resulting from the microscopic collisional dynamics and being associated with the spin-variance in the corresponding collective state. The assumption of low spin polarization and subsequently the linearity (state-independence) of Eqs.~\eqref{eq:DiffEqA} and \eqref{eq:DiffEqB} imply that the mean collective angular momentum variables will be always close to zero. Accordingly, $Q$ can be also assumed to be state-independent. This assumption will be further justified later in the paper, where only the completely unpolarized state is examined. As a consequence of the linearity of the stochastic differential equation, the mean of $\mathbf{\hat{X}} (t)$  follows the equation $d \langle \mathbf{\hat{X}} (t) \rangle =  A \langle \mathbf{\hat{X}} (t) \rangle dt$ with solution 
\begin{equation}
\label{eq:AverageSpinDynamics}
\langle \mathbf{\hat{X}} (t) \rangle = e^{A \tau} \langle \mathbf{\hat{X}} (0) \rangle. 
\end{equation}
To describe the statistical properties of spin-noise we will make extensive use of two-time correlation functions: for observables $\hat{A}$ and $\hat{B}$, the covariance is $\mathcal{R}_{\hat{A},\hat{B}}(\tau) \equiv \langle \hat{A}(\tau) \hat{B}(0) + \hat{B}(0)\hat{A}(\tau) \rangle/2= \mathrm{Re}[\langle \hat{A}(\tau) \hat{B}(0) \rangle]$ since $\hat{A}$ and $\hat{B}$ are Hermitian operators and $(\hat{A} \hat{B})^{\dagger}=\hat{B}^{\dagger}\hat{A}^{\dagger}=\hat{B}\hat{A}$. Additionally, given a column vector of observables $\mathbf{\hat{V}}(t) = (\hat{A},\hat{B},\hat{C},...)^T$ with stationary statistical properties, the covariance matrix is 
\begin{equation}
\begin{split}
\mathcal{R}_{\mathbf{\hat{V}},\mathbf{\hat{V}}}(\tau) &= \frac{1}{2}\Big \{ \langle \mathbf{\hat{V}}(\tau)\mathbf{\hat{V}}^{T}(0)\rangle + [\langle \mathbf{\hat{V}}(0)\mathbf{\hat{V}}^{T}(\tau)\rangle]^{T} \Big \}\\
&= \mathrm{Re}[\langle \mathbf{\hat{V}}(\tau)\mathbf{\hat{V}}^{T}(0)\rangle].
\label{eq:RDef}
\end{split}
\end{equation}  
For commuting or classical observables $\mathbf{V}(t)$, this is simply 
$\mathcal{R}_{\mathbf{{V}},\mathbf{{V}}}(\tau) \equiv \langle \mathbf{{V}}(\tau)\mathbf{{V}}^{T}(0)\rangle$. Here, $\langle \cdot \rangle $ denotes the average over different and independent realizations of the acquisition process. As usual, we assume that the system has equillibrated, or equivalenty that the acquisition time is much longer than $T_1$ \cite{gardiner2009stochastic}. In this work we concentrate on relaxation due to SE collisions. Other relaxations mechanisms, like wall collisions, relaxation due to magnetic gradients or power broadening can be also included in Eq.\eqref{eq:density-matrix} \cite{SavukovRomalis}, but to make evident the SE effects, we assume these are negligible.

A simple formula relates the dynamics matrix $A$ with the noise-strength matrix $Q$ and the steady-state covariance $\mathcal{R}_{\mathbf{\hat{X}},\mathbf{\hat{X}}}(0)$ \cite{gardiner2009stochastic}
\begin{equation}
\label{eq:FluctuationDissipation}
A \mathcal{R}_{\mathbf{\hat{X}},\mathbf{\hat{X}}}(0)+\mathcal{R}_{\mathbf{\hat{X}},\mathbf{\hat{X}}}(0) A^{T} =QQ^{T}.
\end{equation}

We calculate  $\mathcal{R}_{\mathbf{\hat{X}},\mathbf{\hat{X}}}(0)$ in \autoref{sec:Correlations}. 

\subsection{Intrinsic versus thermal spin-noise}

In quantum-limited probing of atomic spin systems, intrinsic spin-noise and the effect of measurement on the spin distribution can play an important role \cite{vasilakis2011stroboscopic, VasilakisPolzik2015, Troullinou}. Here by intrinsic we refer to the noise induced by the uncertainty relation on the interrogated angular momentum components. The magnitude of these effects can be understood by considering the spin uncertainty relations 
\begin{equation}
\label{eq:SpinUncertRel}
\delta F_i \delta F_j \ge \frac{1}{2} |\langle [\hat{F}_i, \hat{F}_j]\rangle| = \frac{1}{2} | \langle i \hat{F}_k\rangle|   =\frac{1}{2} |\langle  \hat{F}_k\rangle|,
\end{equation}
for mutually-orthogonal directions $\{i,j,k\}$. For example, fully-polarized spin-$F$ states with $\langle \hat{F}_z \rangle = \NAtoms f$ have $\delta F_x  \delta F_y = \NAtoms f/2$, which saturates the uncertainty relation. For such polarized states, processes that reduce the uncertainty of one component, e.g. measurement or optical pumping, are required to introduce noise into other components, to satisfy the uncertainty relation \cite{optical_mag}. 

Unpolarized and weakly polarized states also obey Eq.\eqref{eq:SpinUncertRel}, but they satisfy it in a different way. Unpolarized states have uncertainty products $\delta F_i \delta F_j  \approx \NAtoms f(f+1)/3$, comparable to those of polarized states.  A much smaller fraction of their spin variance is imposed by the uncertainty relation, however; for unpolarized states, the right-hand side of Eq.\eqref{eq:SpinUncertRel} is $|\langle \hat{F}_k \rangle| \le \langle \hat{F}_k^2 \rangle^{1/2} \sim \NAtoms ^{1/2} \ll \NAtoms$. The uncertainty relation thus imposes an additional variance that is negligible when compared to the thermal variance of the state. For this reason, the collective angular momentum can be accurately modeled as a classical stochastic variable. The quantum nature of the spin is nonetheless reflected in the structure of the dynamical equations, i.e., in ~Eqs. \eqref{eq:density-matrix},  \eqref{eq:DiffEqA}, and \eqref{eq:DiffEqB}. We note that, because we are dealing with unpolarized states, it suffices to describe the thermal noise contribution. This contribution reflects the thermal state variance, and thus the spin structure of the atomic state, but it does not derive from Eq.\eqref{eq:SpinUncertRel}. More extensive discussions of the effects of measurement back-action on unpolarized states can be found in \cite{toth2010singlet, BehboodPRL2013, BehboodPRL2014, MitchellNatureCom}.

\subsection{Faraday signal}

The dispersive interaction of a linearly polarized weak probe beam with a hot atomic vapor results in  paramagnetic Faraday rotation \cite{PhysRev.163.12,optical_mag} 
by an angle $\hat{\phi}(t)\approx (4 \pi^2 n l\nu/c) \mathrm{Re}[ \hat{a}^{1}(\nu,t)]$, where $n$ is the alkali-metal number density in the vapor cell of length $l$, $\nu$ is the frequency of the monochromatic laser beam interacting with the atomic medium, $c$ is the speed of light and $\hat{a}^{1}(\nu,t)$ is the rank-1 irreducible vector component of the ground-state polarizability tensor $\hat{a}$. In writing the rotation angle, we have considered a detuning much larger than the hyperfine splitting in the excited electronic state such that the contribution of the rank-2 component of the polarizability tensor is negligible. Additionally, we assume that the pressure broadening due to the interaction of the alkali-metal atoms with the atoms of a buffer gas is larger than the Doppler broadening of the optical transition. Under these assumptions, the Faraday rotation takes the form \cite{MitchellNatureCom} \footnote{For a derivation of the rotation angle formula see for example, supplementary material of \cite{PhysRevLett.120.040503}.}
\begin{equation}
\hat{\phi} (t) \approx g_b \hat{F}_{z}^{b}(t)-g_{a}\hat{F}_{z}^{a} (t),
\label{eq:Faraday Rotation}
\end{equation}
with $\hat{F}_{z}^{\alpha}=\sum_{i=1}^{\NAtoms } \hat{f}_{i, z}^{\alpha}$, $\alpha \in \{ a,b \}$ being the collective ground-state angular momentum operator expressed as a uniformly weighted sum over all the single-atom angular momentum operators probed by the laser beam. To avoid labeling complexity, we assume that the probe beam covers the entire vapor cell. The uniform weighting represents a homogeneous atom-light coupling \cite{PhysRevA.89.033850}, while the probe beam is assumed to propagate along the $\hat{\mathbf{z}}$-direction. Here, $a=I+1/2$ and $b=I-1/2$ indicate the upper and lower hyperfine manifolds, respectively, with $I$ being the nuclear spin.  In Eq.\eqref{eq:Faraday Rotation} we ignore the terms of the Faraday signal that oscillate at the hyperfine frequency since they significantly exceed the bandwidth of the usual photodetectors, and they are irrelevant to this work. The coupling constants $g_{\alpha}$ with $\alpha \in \{a,b \}$ are given by
\begin{equation}
g_{\alpha} = \frac{1}{2I+1} \frac{c r_e f_{\mathrm{osc}}}{A_\mathrm{{eff}}} \frac{\nu-\nu_\alpha}{(\nu-\nu_\alpha)^2+(\frac{\Delta \nu}{2})^2},
\label{eq:gjDef}
\end{equation}
where $r_e=2.83 \times 10^{-13}$ cm is the classical electron radius, $f_{\mathrm{osc}}$ is the oscillator strength associated with the particular optical transition, $2I+1$ is the nuclear spin multiplicity, $A_{\mathrm{eff}}$ is the effective beam area, $\Delta \nu$ is the full width at half-maximum (FWHM) optical linewidth and $ \nu-\nu_{\alpha}$ is the optical detuning of the linearly polarized probe light from the electronic manifold of the excited state. Here, $\nu_{\alpha}$ with $\alpha \in \{ a,b \}$ are the optical resonance frequencies of the two ground-state hyperfine levels. For the magnetic fields of interest in this paper, the Zeeman splittings are small compared to the detuning from the optical transition, hence, they can be safely neglected in the description of the atom-light coupling. 

In Faraday probing, the optical rotation is usually measured with a balanced photodetector. The optical signal reaching the detector can be expressed in the Stokes parameter formalism using the input-output relations \cite{RevModPhys.82.1041,Colangelo2017,colangelo2013quantum}
\begin{equation}
\mathcal{\hat{S}}_y\supout (t) \approx \mathcal{\hat{S}}_y\supin (t) + [g_b \hat{F}_{z}^{b}(t)-g_{a}\hat{F}_{z}^{a} (t)] \mathcal{S}_{x}\supin(t), \label{eq:polarimeter}
\end{equation}
where $\mathcal{\hat{S}}_x \equiv [\hat{n}_{\mathrm{ph}}(x)-\hat{n}_{\mathrm{ph}}(y) ]/2  $ is the difference between the number of photons with linear polarization along the $\mathbf{\hat{x}}$- and $\mathbf{\hat{y}}$- directions, and $\mathcal{\hat{S}}_y \equiv [\hat{n}_{\mathrm{ph}}(+45^{\circ})-\hat{n}_{\mathrm{ph}}(-45^{\circ}) ]/2  $ is the difference between the photon numbers  with linear polarization along the $\pm 45^{\circ}$, with respect to the x-axis. The input beam is linearly polarized along $\mathbf{\hat{x}}$, therefore $\mathcal{\hat{S}}_x\supin$ can be treated as a classical variable with $ \mathcal{S}_{x}\supin \equiv \langle \Phi \rangle/2$. Here $\langle \Phi \rangle$ is the mean photon flux where we additionally assume that $\mathcal{S}_x\supin(t) =\mathcal{S}_x\supin= \langle \Phi \rangle /2$ is on average constant during the acquisition process \cite{Julsgaard2001}. The applicability of Eq.\eqref{eq:polarimeter} is limited to small Faraday rotation angles. For unpolarized or low-polarized vapors, the small-angle approximation is satisfied. Finally, due to the off-resonance probing considered here, the effects of optical absorption are negligible and therefore not accounted in Eq.\eqref{eq:polarimeter}.

\section{Spin correlation functions}
\label{sec:Correlations}

\subsection{Steady-state many-body density matrix}

As described in Sec.~\ref{sec:StochasticBlochEquations}, the noise matrix $Q$ can be computed, given the equal-time covariance matrix $\mathcal{R}_{\hat{\mathbf{X}},\hat{\mathbf{X}}}(0)$ and the known dynamical matrix $A$. Physically, the combined noise and relaxation in the stochastic Bloch equations must produce covariances equal to those found in the equilibrium many-body state $\tilde{\rho}$. 

We note that Eqs.~(\ref{eq:DiffEqA}) and (\ref{eq:DiffEqB}) describe the spin evolution ``in the dark,'' i.e., without measurement or back-action effects produced by the probe beam. The relevant equilibrium state is thus that produced by the following process: unitary evolution under the Larmor and hyperfine effects, interrupted at random intervals by sudden spin-exchange collisions between random pairs of atoms within the ensemble. These collisions cause mutual precession of the colliding atoms' electron spins by a random angle.  This process can produce any possible spin state, and, assuming the random parameters do not themselves depend on the spin state, will in time produce each possible state with equal probability. That is, the equilibrium many-body state is a fully-mixed state, which can be written as the product state $\tilde{\rho}_\mathrm{eq} \equiv \rho_\mathrm{th}^{\otimes \NAtoms}$, where $\rho_\mathrm{th} \equiv \mathbbm{1}/d$ is the single-atom thermal state, and $d = (2a +1) + (2b +1)$ is the dimension of $\rho$.

\subsection{Equal-time correlations}
\label{sec:EqualTimeCorrelations}

Given $\tilde{\rho}_\mathrm{eq}$ as just described and  Eq.\eqref{eq:RDef}, we compute the equal-time correlators
\begin{eqnarray}
\mathcal{R}_{\hat{F}_i^{\alpha},\hat{F}_j^{\beta}}(0) &=& \frac{1}{2} \left\langle \hat{F}_i^{\alpha}(0)\hat{F}_j^{\beta}(0) + 
 \hat{F}_j^{\beta}(0)\hat{F}_i^{\alpha}(0)\right\rangle
\nonumber \\ & = & 
\delta_{ij} \delta_{\alpha \beta} \frac{f^{\alpha}(f^{\alpha}+1)(2f^{\alpha}+1)}{6 (2I+1) }\NAtoms.
\label{eq:variance}
\end{eqnarray}
This describes vanishing correlations among different components and different hyperfine levels at $\tau=0$. It is nonetheless compatible with non-vanishing off-diagonal unequal-time correlations, to be computed in the following section. The spectral features of the resulting cross-correlations are similar to what has been shown for heterogeneous spin systems previously \cite{roy2015cross,katz2015coherent}. For future reference we define $\var{(F^{\alpha})} \equiv f^{\alpha} (f^{\alpha}+1) (2f^{\alpha}+1) \NAtoms / [6(2I+1)]$ being the variance of the angular momentum vector in the unpolarized state.

The equal-time correlations described by Eq.~(\ref{eq:variance}) are simple and without structure. For example, there are no equal-time cross-correlations, neither between the different hyperfine components of  a single atom, nor between the spin operators of different atoms. This does not imply these components are truly  uncorrelated; rather, it reflects the fact that the phase of any dynamical behavior involving these components is random. When averaged over this  phase, the equal-time correlations vanish. It is nonetheless possible to observe dynamics involving these distinct components: they are visible in the unequal-time correlations that we now proceed to calculate.

\subsection{Unequal-time correlations}
\label{sec:Unequal-time correlations}

For $\tau>0$, the $\tau$-derivative of $\langle \mathbf{\hat{X}}(\tau)\mathbf{\hat{X}}^{T}(0)\rangle$ is given by
\begin{eqnarray}
\frac{d}{d\tau} \langle \mathbf{\hat{X}}(\tau)\mathbf{\hat{X}}^{T}(0) \rangle 
&=& \langle \frac{A \mathbf{\hat{X}} (\tau)d\tau+Q d {\cal{\mathbf{\hat{W}}}}(\tau)}{d\tau}\mathbf{\hat{X}}^{T}(0)\rangle \nonumber \\
&=&A  \langle \mathbf{\hat{X}}(\tau)\mathbf{\hat{X}}^{T}(0) \rangle,
\end{eqnarray}
where in the second step we used the properties of the Wiener increment assuming there is no correlation between $d {\cal{\mathbf{\hat{W}}}}(\tau)$ and the initial state vector $\mathbf{\hat{X}}^{T}(0)$. Consequently, the $\tau$-dependence of the steady-state covariance matrix is given by
\begin{equation}
  \mathcal{R}_{\mathbf{\hat{X}},\mathbf{\hat{X}}}(\tau) =
  \begin{cases}
e^{A\tau}\mathcal{R}_{\mathbf{\hat{X}},\mathbf{\hat{X}}}(0) & ,\tau > 0\\
\\
\mathcal{R}_{\mathbf{\hat{X}},\mathbf{\hat{X}}}(0) e^{-A^{T}\tau} & ,\tau < 0
 \end{cases} \label{eq:cov}
\end{equation}
This is an important result, indicating that the $\tau$ evolution of the correlation $\mathcal{R}_{\mathbf{\hat{X}},\mathbf{\hat{X}}}(\tau)$  is governed by the same equation as the time evolution of the mean $ \langle \mathbf{\hat{X}} (t) \rangle$, c.f. Eq.\eqref{eq:AverageSpinDynamics}. This result is known in the literature as the regression theorem \cite{gardiner2009stochastic}.  Hereafter, we concentrate only on correlations for $\tau>0$, whilst equivalent results can be obtained for $\tau<0$. In Eq.\eqref{eq:cov}, $A$ is the $6 \times 6$ matrix
\begin{equation}
\label{eq:ADef}
 \setlength{\arraycolsep}{0.1pt}
\small{A=} 
\begin{pmatrix} 
-R_{\mathrm{se}} \kappa_{aa}  & 0 & -\omega_0 & R_{\mathrm{se}} \kappa_{bb} & 0 & 0 \\
0& -R_{\mathrm{se}} \kappa_{aa} & 0& 0 &
R_{\mathrm{se}} \kappa_{bb} & 0
\\
\omega_0 & 0 & -R_{\mathrm{se}} \kappa_{aa} & 0 & 0 &  R_{\mathrm{se}} \kappa_{bb} \\ 
R_{\mathrm{se}} \kappa_{aa} & 0 & 0 &  -R_{\mathrm{se}} \kappa_{bb} & 0 & \omega_0 \\
0 & R_{\mathrm{se}} \kappa_{aa} & 0 & 0 & -R_{\mathrm{se}} \kappa_{bb} & 0 \\
0 & 0 & R_{\mathrm{se}} \kappa_{aa} & -\omega_0 & 0 & -R_{\mathrm{se}} \kappa_{bb} 
\end{pmatrix} .
\end{equation}

Analytical expressions for the covariance components can be derived by eigen-decomposing the matrix exponential into $e^{A \tau}= V e^{\Lambda \tau} V^{-1}$. The details of the eigenspectrum of $A$ are presented in Appendix \ref{sec:Eigenspectrum}. In the presence of the magnetic field along the $\hat{\mathbf{y}}$-direction, the dynamics are separated into longitudinal (i.e., along the B-field direction) and transverse, with vanishing correlations between the two. Simple analytical expressions are obtained for both components. The longitudinal covariances are given by
\begin{equation}
\mathcal{R}_{\hat{F}_y^{a},\hat{F}_y^{a}}(\tau) = \frac{\var{(F^{a})}}{\kappa_{aa}+\kappa_{bb}}\Big[\kappa_{bb}+\kappa_{aa}e^{-(\kappa_{aa}+\kappa_{bb})R_{\mathrm{se}}\tau} \Big],
\label{eq:longitudinal1}
\end{equation}
\begin{equation}
\mathcal{R}_{\hat{F}_y^{b},\hat{F}_y^{b}}(\tau) = \frac{\var{(F^{b})}}{\kappa_{aa}+\kappa_{bb}}\Big[\kappa_{aa}+\kappa_{bb}e^{-(\kappa_{aa}+\kappa_{bb})R_{\mathrm{se}}\tau} \Big],
\end{equation}
\begin{equation}
\mathcal{R}_{\hat{F}_y^{a},\hat{F}_y^{b}}(\tau) = \frac{\var{(F^{b})}}{\kappa_{aa}+\kappa_{bb}}\kappa_{bb}\Big[1-e^{-(\kappa_{aa}+\kappa_{bb})R_{\mathrm{se}}\tau} \Big], 
\end{equation}
\begin{equation}
\mathcal{R}_{\hat{F}_y^{b},\hat{F}_y^{a}}(\tau) = \frac{\var{(F^{a})}}{\kappa_{aa}+\kappa_{bb}}\kappa_{aa}\Big[1-e^{-(\kappa_{aa}+\kappa_{bb})R_{\mathrm{se}}\tau} \Big]. 
\label{eq:longitudinal4}
\end{equation}
We note that $\kappa_{aa} \var{(F^a)} = \kappa_{bb} \var{(F^b)}$ and thus, $\mathcal{R}_{\hat{F}_y^{a},\hat{F}_y^{b}}(\tau) = \mathcal{R}_{\hat{F}_y^{b},\hat{F}_y^{a}}(\tau) $. The transverse covariance components can be summarized in a simple formula
\begin{equation}
\mathcal{R}_{\hat{F}_i^{\alpha},\hat{F}_j^{\beta}}(\tau) = 2 \Big( \mathrm{Re}[c_1 e^{-(\Gamma_{-}+i \Omega_{-})\tau}]+
 \mathrm{Re}[c_2 e^{-(\Gamma_{+}+i \Omega_{+})\tau}] \Big), \label{eq:cov-tau}
\end{equation}
where the coefficients $c_1 \equiv c_1(F^{\alpha}_{i},F^{\beta}_{j})$ and $ c_2 \equiv c_2(F^{\alpha}_{i},F^{\beta}_{j})$ are different for each covariance component and are discussed in Appendix \ref{sec:time evolution}. For slow spin-exchange ($\omega_0 \gg R_{\mathrm{se}}$), the relaxation rates $\Gamma_{\pm}$ entering Eq.\eqref{eq:cov-tau} are given by $\Gamma_{-} \approx \kappa_{aa}R_{\mathrm{se}}$ and $\Gamma_{+} \approx \kappa_{bb}R_{\mathrm{se}}$ and are of comparable orders of magnitude. Additionally, the two modes $\Omega_{\pm}$ precess at the same frequency, in opposite directions, and with independent phases; therefore, they are related through $\Omega_{+}= -\Omega_{-}$. These approach $\pm \omega_0$ when $\omega_0 \gg R_{\mathrm{se}}$, but can differ significantly from it in the rapid SE regime ($\omega_0 \ll R_{\mathrm{se}}$) where the precession frequency is shifted down to
\begin{equation}
\Omega_+ = - \Omega_{-} \approx \frac{3(2I+1)}{3+4I(I+1)}\omega_0.
\end{equation} 
The corresponding relaxation rates are approximately given by \cite{PhysRevA.16.1877}
\begin{equation}
\Gamma_{+} \approx(\kappa_{aa}+\kappa_{bb})R_{\mathrm{se}},
\end{equation}
\begin{equation}
\Gamma_{-} \approx \frac{\omega_0^2}{R_{\mathrm{se}}} \frac{2 I(I+1)(2I-1)(2I+3)}{(4I^2 + 4I +3)}.
\end{equation}

\begin{figure}
\centering
\includegraphics[width=7.5cm]{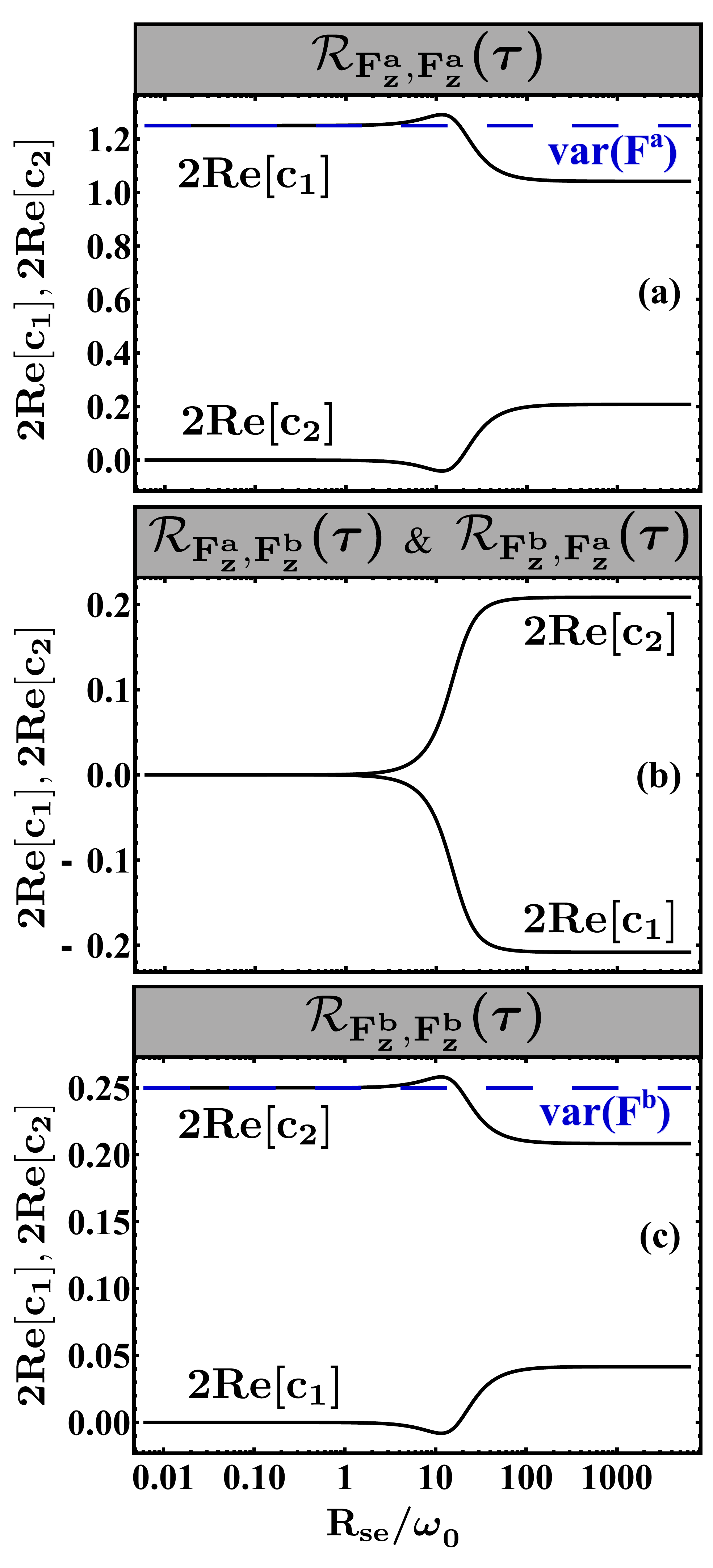} 
\caption{Two times the real parts of the coefficients $c_1(F^{\alpha}_{z},F^{\beta}_{z})$ and $c_2(F^{\alpha}_{z},F^{\beta}_{z})$ entering the time-evolution of the transverse covariance components (a) $\mathcal{R}_{F_z^{a},F_z^{a}}(\tau)$, (b) $\mathcal{R}_{F_z^{a},F_z^{b}}(\tau) $ and $\mathcal{R}_{F_z^{b},F_z^{a}}(\tau)$, and (c) $ \mathcal{R}_{F_z^{b},F_z^{b}}(\tau)$, according to Eq.\eqref{eq:cov-tau}. The single-atom variances in the two hyperfine multiplets are shown with blue dashed lines. The plot corresponds to a $^{87}$Rb vapor at a magnetic field of $B=10$ mG.}
\label{fig:cov_coefficients}
\end{figure} 

In \autoref{fig:cov_coefficients} the real parts of $c_1$ and $c_2$ are plotted against $R_{\mathrm{se}}/ \omega_0$ while in Appendix \ref{sec:time evolution} analytical expressions for the coefficients are presented for the covariance components along the $\mathbf{\hat{z}}$-direction. 

\begin{figure*}[htp]
\centering
\includegraphics[width=18cm]{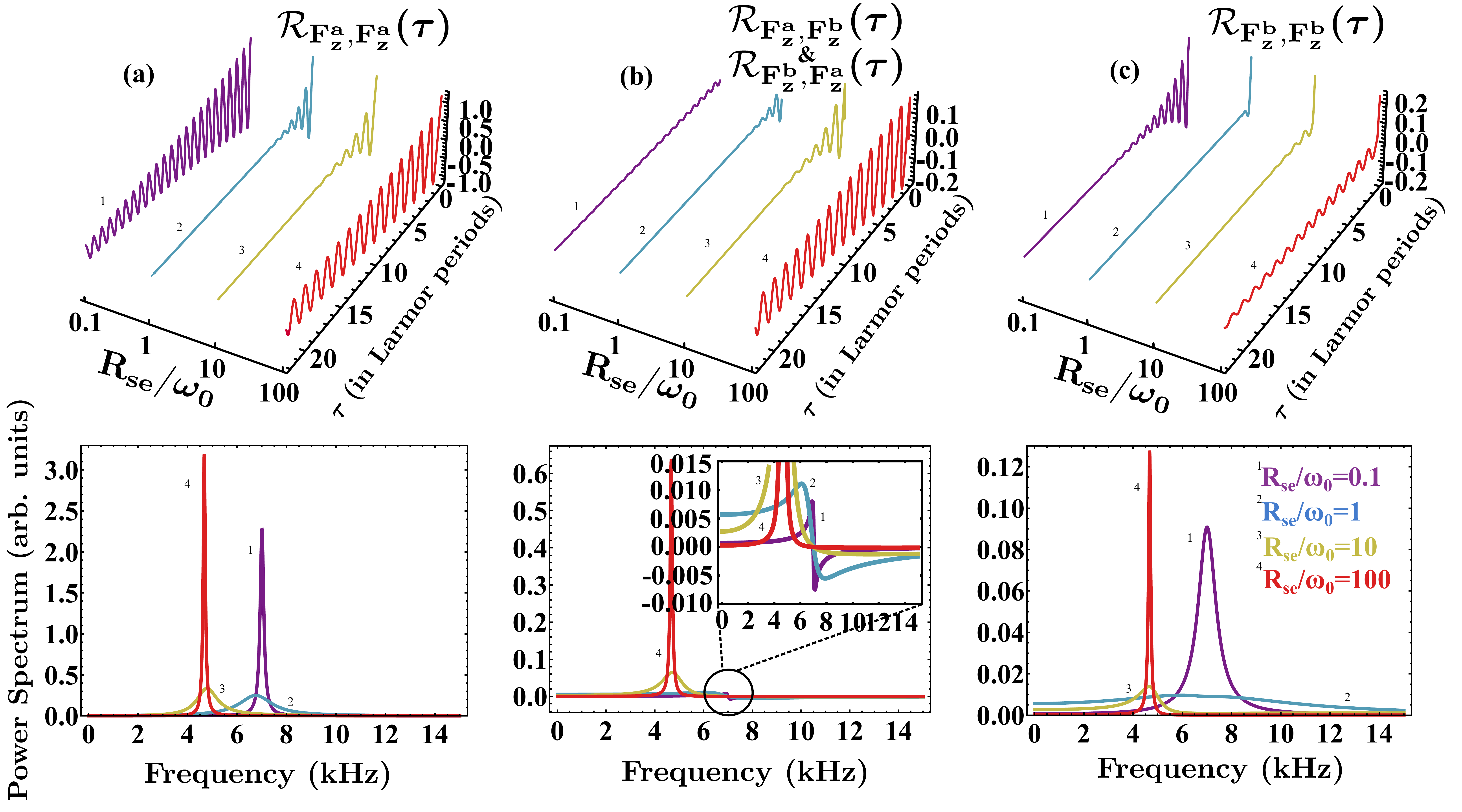} 
\caption{$\tau$-dependence  of the steady-state covariance components  probed by a linearly polarized electric field (top row) and the corresponding power spectra (bottom row) for a $^{87}$Rb vapor at four different values of the spin-exchange rate: $R_{\mathrm{se}}/ \omega_0=0.1$ (no.1-purple), $1$ (no.2-blue), $10$ (no.3-yellow) and $100$ (no.4-red). The magnitude of the magnetic field is fixed to $B=10$ mG corresponding to a  Larmor frequency of $\omega_0 \approx 2 \pi \times 7.0$ \si{\kilo\hertz} and a Larmor period of $\tau_{_L}\approx 0.1$ \si{\milli\second}. In the time domain, an increasing coherence time for all covariance components as approaching the SERF regime is illustrated. On top of that, the cross-covariances in (b) reveal an increasing amplitude, indicating strong cross-correlations between the two hyperfine levels. The power spectra behave accordingly. In (a) and (c) the spectrum is given by a single Lorentzian centered at the corresponding precession frequency, while the spectrum in (b) is of dispersive form that tends towards a Lorentzian in the SERF regime. 
For rapid SE, the Larmor frequency is slown-down to the value $|\Omega_+|=|\Omega_-| \approx 3(2I+1)\omega_0/[3+4I(I+1)]\approx 2 \pi \times 4.7$ \si{\kilo\hertz}, also the linewidth is approximately given by $\Gamma_{-} \approx \frac{\omega_0^2}{R_{\mathrm{se}}} [2 I(I+1)(2I-1)(2I+3)]/(4I^2 + 4I +3)$, both constituting strong footprints of the SE interaction on the spin-noise spectrum. The spectral lines in the lower manifold are broader due to the large contribution of the fast relaxation rate $\Gamma_{+}= (\kappa_{aa}+\kappa_{bb})R_{\mathrm{se}}$, affecting the wings of the resonance.}
\label{fig:time-cov}
\end{figure*} 

At high magnetic fields the fast spin precession ``averages out'' correlations between the two oppositely precessing states. In contrast, the dynamics in the two hyperfine levels are interlocked by the rapid SE collisions as the spins precess slowly in the magnetic field \cite{PhysRevA.16.1877}; thus, enabling unequal-time cross-correlations to build-up. These effects are imprinted on the behavior of  $c_1(F^{\alpha}_{i},F^{\beta}_{j})$ and $ c_2(F^{\alpha}_{i},F^{\beta}_{j})$ as illustrated in \autoref{fig:cov_coefficients}b). Both coefficients transition from a zero value at high magnetic fields to an amplitude roughly equal to that of $\mathcal{R}_{\hat{F}_z^{b},\hat{F}_z^{b}}(\tau)$, shown in \autoref{fig:cov_coefficients}c). Further, since the absolute values of the two coefficients are equal, according to Eq.\eqref{eq:cov-tau}, the relaxation rates $\Gamma_+$ and $\Gamma_-$ contribute equally in the $\tau$-evolution. In contrast, as seen in \autoref{fig:cov_coefficients}a) and c), the dynamics of  $\mathcal{R}_{\hat{F}_z^{a},\hat{F}_z^{a}}(\tau)$ and $\mathcal{R}_{\hat{F}_z^{b},\hat{F}_z^{b}}(\tau)$ are mostly dominated by a single eigenmode with relaxation rate $\Gamma_{-}$ and $\Gamma_{+}$, respectively. As a consequence, the two auto-correlations relax differently.

Apart from the correlated dynamics, \autoref{fig:cov_coefficients} indicates also the way spin-noise power is distributed among the two transverse modes appearing in Eq.\eqref{eq:cov-tau} and subsequently across the frequency spectrum. This is the case since $\mathrm{Re}[c_1]$ and $\mathrm{Re}[c_2]$ directly refer to the spin-noise power associated with the particular mode. In the SERF regime, it is apparent that in the upper hyperfine manifold, spin-noise power is mostly concentrated in a narrow spectral region, in the vicinity of the precession frequency $\Omega_{-}$ since $\lambda_1=\Gamma_{-}+i \Omega_{-}$ dominates the correlations. In contrast, the power in the lower manifold is distributed across a broad range of frequencies around $\Omega_{+}$, reflecting the dominance of $\lambda_2=\Gamma_{+}+i\Omega_{+}$.  Finally, the cross-correlation power is zero as expected according to Eq.\eqref{eq:variance}. 

An overall demonstration of the discussed effects is illustrated in \autoref{fig:time-cov}. Both the time and the frequency domain of the covariance functions are plotted across the SE-SERF transition while the magnetic field is kept constant at $B=10$ mG. When the SE rate starts to exceed the Larmor frequency, $\omega_0 \approx 2 \pi \times 7$ \si{\kilo \hertz}, an increasing coherence time and a frequency shift is demonstrated for all covariance components. In \autoref{fig:time-cov}a) and c) the auto-covariance functions $\mathcal{R}_{F_z^{a},F_z^{a}}(\tau)$ and $\mathcal{R}_{F_z^{b},F_z^{b}}(\tau)$ in the upper and the lower hyperfine manifolds, respectively, are presented together with the corresponding power spectra. The spectral lines in the lower manifold are broad due to the large contribution of the fast relaxation
rate $\Gamma_{+}= (\kappa_{aa}+\kappa_{bb})R_{\mathrm{se}}$, affecting the wings of the resonance. In contrast, narrow resonant features with considerable amplitudes are obtained for the upper manifold consistent with \autoref{fig:cov_coefficients}a). Finally, in \autoref{fig:time-cov}b) the cross-correlation dynamics are illustrated. Details about the spectral features as well as implications of the peculiar lineshapes are discussed in the following sections.

\subsection{Spin-noise spectra}

The power spectral density is the Fourier transform of the auto-covariance function of the detected optical signal \cite{PhysRevA.96.062702}
\begin{equation}
S(\nu) =  \int_{-\infty}^{+ \infty} \mathcal{R}_{\mathcal{\hat{S}}_y\supout,\mathcal{\hat{S}}_y\supout}(\tau) e^{-2 \pi i  \nu \tau} d \tau. \label{eq:PSDmain}
\end{equation}
Plugging Eq.\eqref{eq:polarimeter} into Eq.\eqref{eq:PSDmain} and given that the Stokes and the angular momentum operators are Hermitian, it is apparent that three distinct terms contribute to the covariance function of the Faraday signal: i) a white photon shot noise originating from $\mathcal{\hat{S}}_y\supin(t)$, ii) the atomic covariance indicating correlations of the atomic signal between different times and iii) the cross-terms corresponding to correlations between the atomic angular momentum and the fluctuations of the input light $\mathcal{\hat{S}}_y\supin(t)$. The photon shot-noise auto-covariance function for a coherent field is given by $\mathcal{R}_{\mathcal{\hat{S}}_y\supin,\mathcal{\hat{S}}_y\supin}(\tau) = \mathrm{Re}[\langle \mathcal{\hat{S}}_y\supin (\tau) \mathcal{\hat{S}}_y\supin (0) \rangle] =  \langle \Phi \rangle \delta(\tau)$ with $\delta(\tau)$ being the Dirac delta function. Hereafter, we assume that polarization fluctuations of the input light are not affecting the atomic variables and therefore they are not correlated with the spin-evolution i.e., the cross-terms are not contributing to the spin-noise spectrum. For an unpolarized vapor probed by an off-resonance beam, the backaction effect is negligible \cite{optical_mag}. The resulting auto-covariance function of the detected optical signal can be therefore expressed as

\begin{widetext}
\begin{equation}
\mathcal{R}_{\mathcal{\hat{S}}_y\supout,\mathcal{\hat{S}}_y\supout}(\tau) \approx  \langle \Phi \rangle \delta(\tau)  + \frac{\langle\Phi\rangle^2}{4} \left\{
g_a^2 \mathcal{R}_{\hat{F}_z^a,\hat{F}_z^a}(\tau)  + g_b^2  \mathcal{R}_{\hat{F}_z^b,\hat{F}_z^b}(\tau) - g_a g_b [\mathcal{R}_{\hat{F}_z^a,\hat{F}_z^b}(\tau) + \mathcal{R}_{\hat{F}_z^b,\hat{F}_z^a}(\tau) ] 
\right\}.
\label{Eq:auto-corr}
\end{equation}
\end{widetext}

In order to obtain the power spectrum given by Eq.\eqref{eq:PSDmain} and Eq.\eqref{Eq:auto-corr}, we calculate the Fourier transform of each of the covariance components as demonstrated in the previous section
\begin{equation}
\begin{split}
 S_{\mathbf{\hat{X}},\mathbf{\hat{X}}}(\omega)  &=\int_{-\infty}^{0}  \mathcal{R}_{\mathbf{\hat{X}},\mathbf{\hat{X}}}(0) e^{-A^{T}\tau}  e^{i\omega \tau} d \tau\\ 
&+ \int_{0}^{+\infty} e^{A\tau}\mathcal{R}_{\mathbf{\hat{X}},\mathbf{\hat{X}}}(0) e^{-i\omega \tau} d \tau , \label{eq:power spec}
\end{split}
\end{equation}
By applying the spectral decomposition $e^{A\tau}=V e^{\Lambda \tau} V^{-1}$ and by explicitly calculating the integral $\int_{0}^{+\infty} e^{[\Lambda-i \omega \mathbb{1}]\tau} d\tau=(\Lambda-i \omega \mathbb{1})^{-1}$ we obtain 
\begin{equation}
 S_{\mathbf{\hat{X}},\mathbf{\hat{X}}}(\omega)  =  
 -V (\Lambda-i\omega \mathbb{1})^{-1} V^{-1} \mathcal{R}_{\mathbf{\hat{X}},\mathbf{\hat{X}}}(0) + \mathrm{c.c}. \label{eq:power spec1}
\end{equation}
The resulting power spectra are real for all covariance components with the longitudinal being centered at zero and the transverse at $\pm \Omega_{\pm}$. Eq.\eqref{eq:power spec1} shows the dependence of the power spectrum on the eigenspectrum $\Lambda$ in the presence of SE collisions, as was first illustrated in \cite{PhysRevA.16.1877}. Both $\Gamma_{\pm}$ are contributing to the broadening of the spectrum, each weighted according to \autoref{fig:cov_coefficients}.

\begin{figure}
\centering
\includegraphics[width=8.0cm]{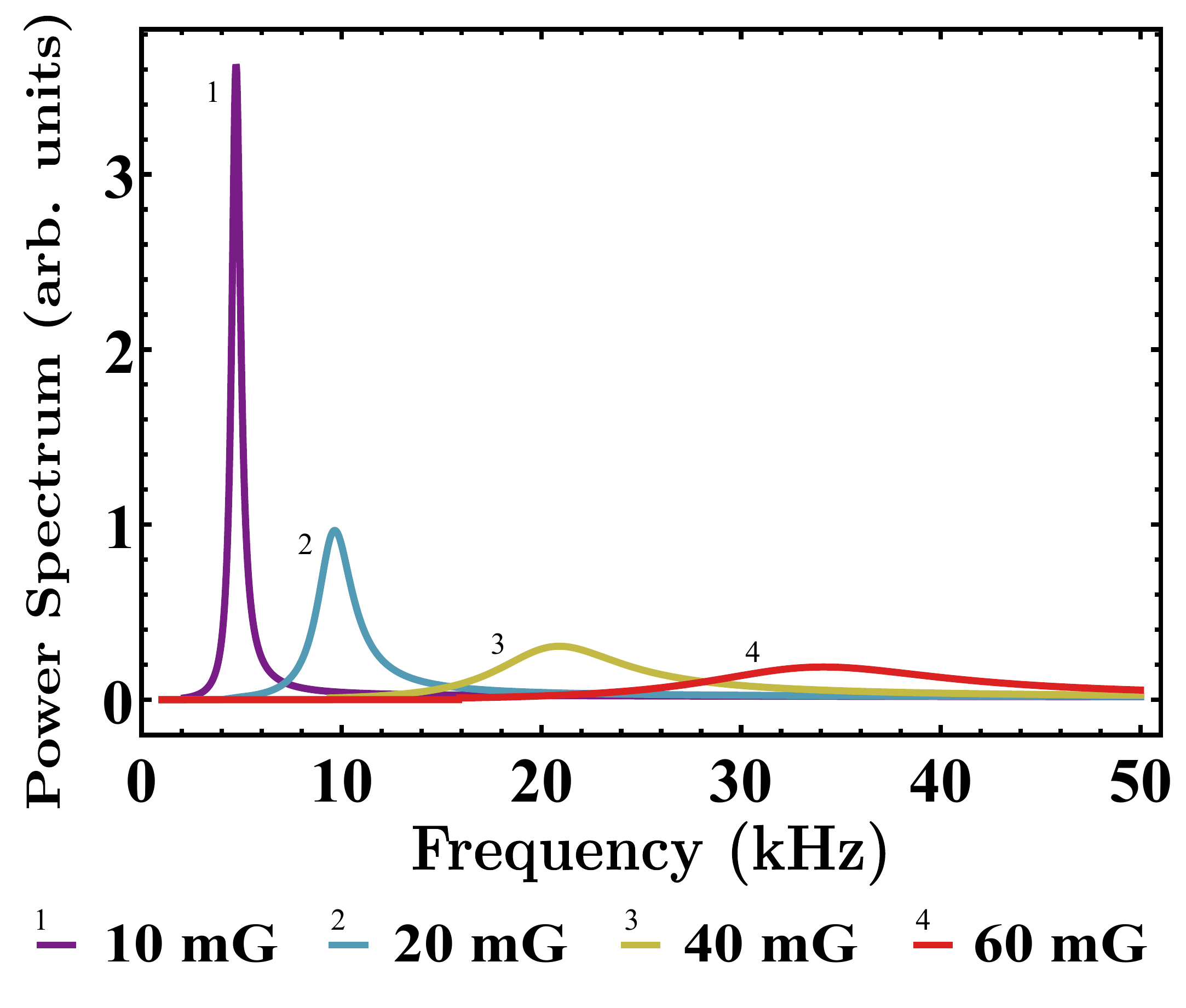} 
\caption{Atomic spin-noise power spectrum for a pure $^{87}$Rb vapor as a function of the magnetic field. The spectrum is calculated at a temperature of $T=200^{\circ}$ C corresponding to an alkali-metal number density of $n_{\mathrm{Rb}}=9.2 \times 10^{14}$ cm$^{-3}$ and consequently to a spin-exchange rate of $R_{\mathrm{se}}\approx 2 \pi \times 134$ \si{\kilo\hertz}. The chosen detuning is $50$ \si{\giga\hertz} blue detuned from the $D_1$ transition and the pressure broadening $\Delta \nu= 10$ \si{\giga\hertz}.}
\label{fig:PSDvsB}
\end{figure}

In \autoref{fig:PSDvsB} we present the atomic power spectrum resulting from the Fourier transform of the atomic covariance components appearing in the expression for the dispersive Faraday rotation signal, Eq.\eqref{Eq:auto-corr}.
We note that the spectrum is qualitatively in agreement with the spectral features reported in \cite{MitchellNatureCom}. It is apparent that the lineshape is dominated by the dynamics in the upper hyperfine multiplet since the amplitude of $\mathcal{R}_{\hat{F}_z^a,\hat{F}_z^a}(\tau)$ greatly exceeds the amplitudes of the remaining components. Although the underlying generation mechanism of the spectrum is the sum of  eight complex Lorentzians appearing in Eq.\eqref{eq:PowerSpec}, in practice, the experimentally relevant lineshape resembles a simple Lorentzian feature centered at $|\Omega_{\pm}|$.  

Due to the intriguing behavior of the correlation functions in the SERF regime, in the next section we show that by appropriately choosing the optical detuning of the Faraday probe we can control the spectral distribution of spin-noise power and under certain circumstances obtain significant reduction as a consequence of the peculiar lineshape.

\section{Predicted phenomena in SNS }
\label{sec:PredictedPhenomena}

We now describe specific observable phenomena predicted by the theory described in the preceding sections. 

\subsection{Probe detuning dependence}
\label{sec:ProbeDetuning}

\begin{figure}[htp]
\centering
\includegraphics[width=8.0cm]{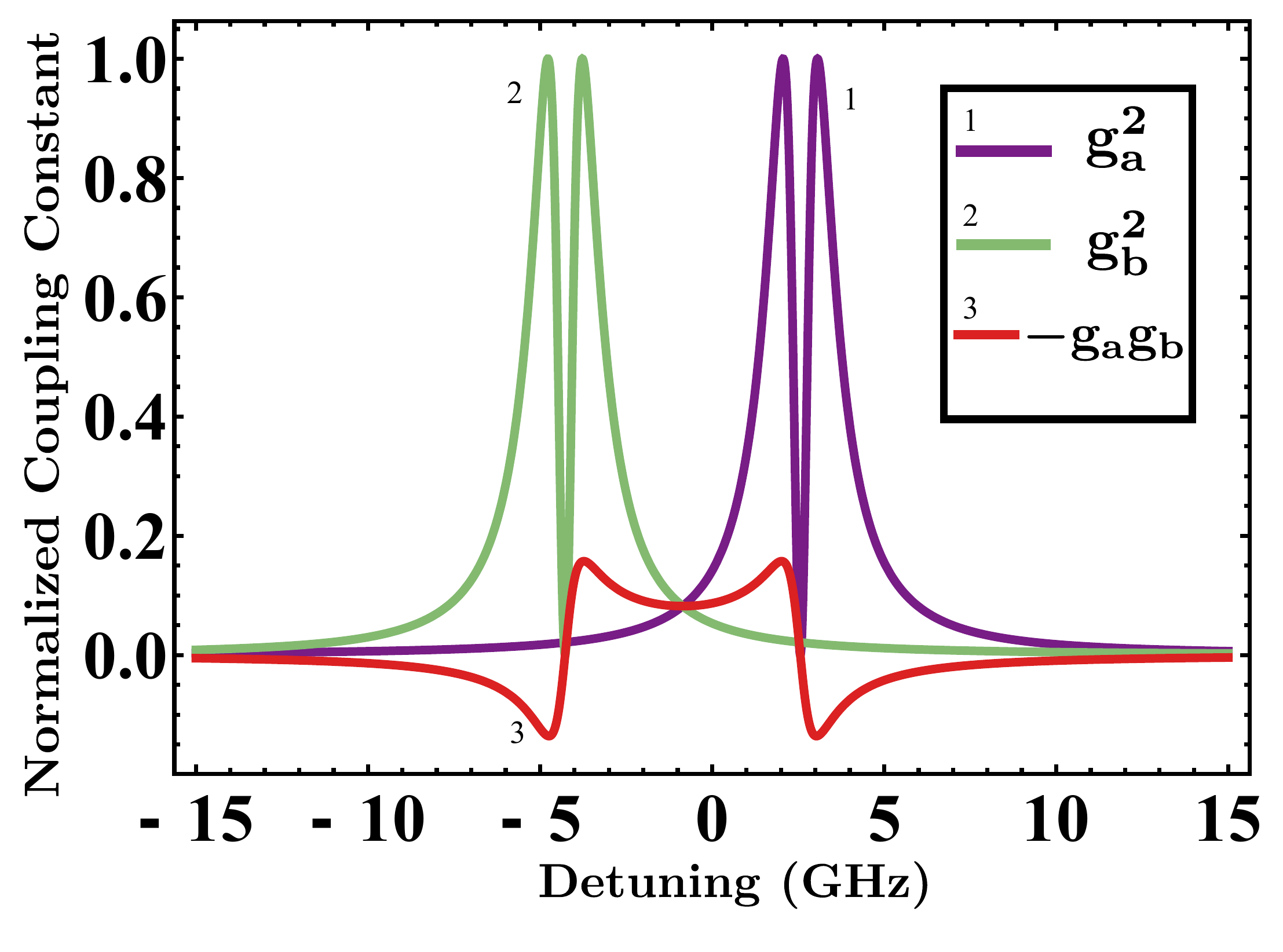} 
\caption{Coupling constants $g_a^2$, $g_b^2$ and $-g_a g_b$ appearing in Eq.\eqref{Eq:auto-corr} as a function of the optical detuning from the $D_1$ transition of $^{87}$Rb. The spectrum corresponds to an optical linewidth of $\Delta \nu=\SI{1}{\giga\hertz}$ and the lineshapes are normalized to the maximum value of $g_a^2$.}
\label{fig:detuning}
\end{figure}

Besides $c_n(F^{\alpha}_{i},F^{\beta}_{j})$ in \autoref{fig:cov_coefficients}, which shape the spectrum and distribute the noise power for a given ratio of the alkali-metal number density and the magnetic field, an equally significant weighting parameter is the optical detuning of the probe. In \autoref{fig:detuning} we plot the coupling constants entering Eq.\eqref{Eq:auto-corr} as a function of the frequency of the probe for a $^{87}$Rb vapor at low buffer gas pressure, corresponding to an optical broadening of $\Delta \nu =1 $ \si{\giga\hertz}. Due to the unique behavior of each covariance component, by suitably choosing the detuning, both the lineshape and the spectral distribution of the spin-noise power can be modified. It is worth noting that in this near-resonant regime optical attenuation of the signal will be important, especially at high vapor densities.

In SERF, as a consequence of the strong cross-correlations between the two hyperfine levels, untypical spin-noise spectra are obtained. For instance, by choosing a red detuning at the wing of the lower hyperfine resonance, we eliminate the contribution of the strong covariance component in the upper hyperfine manifold and consequently the lineshape is dominated only by cross-covariances and covariances in the lower manifold \footnote{A detailed, quantitative treatment of the polarizability Hamiltonian can be found in \cite{PhysRev.163.12}.  Using the equations presented in that reference (Eqs. III.1-III.3), we found that in all of the scenarios studied in the manuscript, higher spin moments affect the atomic susceptibility, and correspondingly the optical readout, at a level of less than $2 \%$.

In principle, the effects of the optical absorption close to the resonance can be kept to be very small, either by increasing the area of the beam or by reducing the photon flux, thus rendering insignificant the perturbation to the spin dynamics due to light-absorption}
According to \autoref{fig:cov_coefficients}, both of the preceding components have comparable amplitudes, and hence, as seen in \autoref{fig:detuning}, by appropriately choosing the detuning we can make their contribution in the Faraday signal equally significant. 

\subsection{Partial cancellation of spin-noise at the precession frequency}
\label{sec:noise cancellation}

\begin{figure}
\centering
\includegraphics[width=7.0cm]{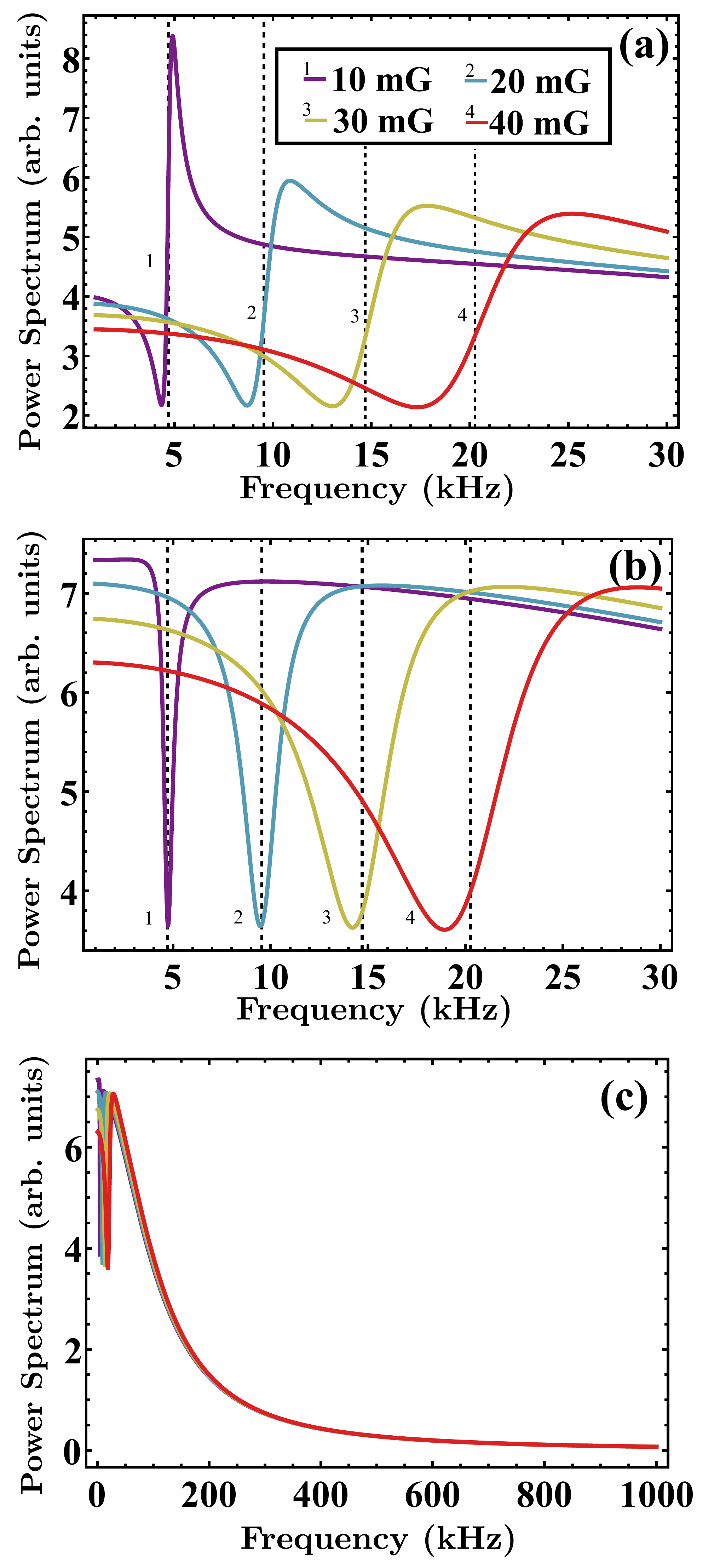} 
\caption{Spin-noise spectrum of $^{87}$Rb for various detunings. (\textbf{a}) The probe laser frequency is tuned to $\nu=-5.7$ \si{\giga\hertz} from the $D_1$ electronic transition, (\textbf{b}) and (\textbf{c}) the laser frequency is $\nu=-6.2$ \si{\giga\hertz}. The optical linewidth is $\Delta \nu=1$ \si{\giga\hertz}. The temperature is $T=200^{\circ}$ C corresponding to a number density of $n_{\mathrm{Rb}}=9.2 \times 10^{-14}$ cm$^{-3} $ and to a spin-exchange rate of $R_{\mathrm{se}} \approx 2 \pi \times 134$ \si{\kilo\hertz}. In (\textbf{c}) we extend the plot of (\textbf{b}) to higher frequencies. The dashed lines correspond to the precession frequencies $|\Omega_{\pm}|$ for each applied magnetic field.}
\label{fig:spectrum_detuning}
\end{figure}

In addition, the opposite sign between the coupling constants $g_{b}^2$ and $-g_{a} g_{b}$ leads to unusual spectral features as illustrated in \autoref{fig:spectrum_detuning}. Resonant dips can be observed at the spin-precession frequency as indicated in \autoref{fig:spectrum_detuning}b) and c). In this case, the spectrum resembles a band-rejection filter centered at the precession frequency with a spectral distribution such that spin-noise power is reduced on resonance. The spectrum is an effect of the coherent cross-correlations that are subtracted from the $\lambda_2$-dominated spectrum in the lower manifold. Finally, we note that in the same regime, fine adjustment of the detuning across the red, near-wing of the lower hyperfine resonance, results in modified spectra such as those, for instance, in  \autoref{fig:spectrum_detuning}a), 
which are significantly different from the Lorentzian lineshape usually encountered in SNS experiments. 

\subsection{Variable integrated spin-noise power}
\label{sec:IntegratedPower}

In many experimental scenarios the spin-noise power of the thermal state is obtained either from the variance of the time-averaged signal in a sample time $\Delta t$, $
\bar{\hat{F}}_i^{\alpha} (t) = \frac{1}{\Delta t} \int_{t}^{t+\Delta t} \hat{F}_i^{\alpha} (t') dt' $, after averaging many repetitions  \cite{PhysRevLett.104.133601,VasilakisPolzik2015}, or by fitting or numerically integrating the steady-state spectrum \cite{katsoprinakis,Dellis,Squeezed-Spin-Noise,MitchellNatureCom}. In a time interval $[t,t+\Delta t]$, each term in Eq.\eqref{Eq:auto-corr} contributes to the total power of the detected Faraday signal. The spin-noise power in that time window is intimately related to the atomic covariance through
\begin{equation}
\begin{split}
\mathcal{P}_{\hat{F}_z^{\alpha},\hat{F}_z^{\beta}} &= \frac{1}{\Delta t^2} \int_{0}^{\Delta t} \tau d\tau \left[ -\mathcal{R}_{\hat{F}_z^{\alpha},\hat{F}_z^{\beta}}(-\tau)-\mathcal{R}_{\hat{F}_z^{\alpha},\hat{F}_z^{\beta}}(\tau) \right]  \\
&+ \frac{1}{\Delta t}  \int_{-\Delta t}^{\Delta t} d\tau \mathcal{R}_{\hat{F}_z^{\alpha},\hat{F}_z^{\beta}}(\tau). \label{eq:CorrectedVarTimeWindow}
\end{split}
\end{equation}
The derivation of the preceding equation is presented in Appendix \ref{sec:MB}. In the frequency domain, the power is obtained by realizing the fact that the correlation function is the Fourier transform of the power spectrum. From Eq.\eqref{eq:CorrectedVarTimeWindow} we find 
\begin{equation}
\mathcal{P}_{\hat{F}_z^{\alpha},\hat{F}_z^{\beta}} =  \frac{1}{2\pi}\int_{-\infty}^{\infty} d\omega S_{\hat{F}_z^{\alpha},\hat{F}_z^{\beta}}(\omega) \left[ \mathrm{sinc}\left(\frac{\omega \Delta t}{2} \right) \right]^2, \label{eq:NoisePowerSpecBandwidth}
\end{equation}
where $\mathrm{sinc}(x)=\sin{(x)}/x$. As $ \Delta t \rightarrow 0 $, from Eq.\eqref{eq:NoisePowerSpecBandwidth} we obtain
\begin{equation}
\begin{split}
\mathcal{P}_{\hat{F}_z^{\alpha},\hat{F}_z^{\beta}} &= \frac{1}{2\pi}\int_{-\infty}^{\infty} d\omega S_{\hat{F}_z^{\alpha},\hat{F}_z^{\beta}}(\omega)\\
&=\frac{1}{2\pi}\int_{-\infty}^{\infty} \int_{-\infty}^{\infty} d\omega \mathcal{R}_{\hat{F}_z^{\alpha},\hat{F}_z^{\beta}}(\tau)e^{i \omega \tau}   d \tau\\
&= \mathcal{R}_{\hat{F}_z^{\alpha},\hat{F}_z^{\beta}}(0),
\end{split}
\end{equation}
where in the last step we used the property of the Dirac delta function $\delta(\tau-\alpha)=\frac{1}{2\pi}\int_{-\infty}^{+\infty} d\omega e^{i \omega (\tau-\alpha)}$. 
Using the steady-state covariance at $\tau=0$, the total power of the polarimeter output is expressed as
\begin{equation}
\mathcal{R}_{\mathcal{\hat{S}}_y\supout,\mathcal{\hat{S}}_y\supout}(0)= \mathrm{PSN} + \frac{\langle\Phi\rangle^2}{4} 
[g_a^2 \var(F^a)  + g_b^2  \var(F^b)],
\label{Eq:SN-power}
\end{equation}
being in agreement with \cite{vasilakis2011stroboscopic,PhysRevLett.106.143601}. Here $\mathrm{PSN}$ is the total photon shot-noise power. As is evident in Eq.\eqref{Eq:SN-power},  although the cross-correlations are not contributing to the total spin-noise power, it is possible experimentally to measure a non-zero contribution due to the limited bandwidth of the acquisition process, defined as $\mathrm{BW}=1/(2 \Delta t)$. As can be seen in the spectrum of \autoref{fig:time-cov}(b), when in the SERF regime, the cross-correlation power, given by the integral under the spectral lineshape, is equally distributed between the resonant peak and the negative far wing that extends to the high-frequency part of the spectrum. Due to the limited bandwidth, frequencies higher than the Nyquist frequency are not entering the signal, thus resulting in a non-zero and positive cross-power. This means that according to Eq.\eqref{eq:NoisePowerSpecBandwidth} and  Eq.\eqref{Eq:auto-corr}, the amount of the remaining non-zero cross-correlation power will be subtracted from the total power of the Faraday signal, resulting in a reduced spin-noise power as opposed to the expected one, given by Eq.\eqref{Eq:SN-power}.

Apart from the non-zero cross-correlation power, we predict an additional and similar mechanism, realized in experimental implementations of SNS, that leads to further reduction of the noise power compared to the level expected from the thermal spin state. In the SERF regime, because of the fast decay $\Gamma_{+}$ that enters both of the auto-correlation functions, the corresponding spectra extend to high frequencies and are therefore constrained by the acquisition. Consequently, the integrated spectrum in Eq.\eqref{eq:NoisePowerSpecBandwidth} contains only a portion of the presumed noise power of the thermal spin state. 

\begin{figure}[htp]
\centering
\includegraphics[width=8.5cm]{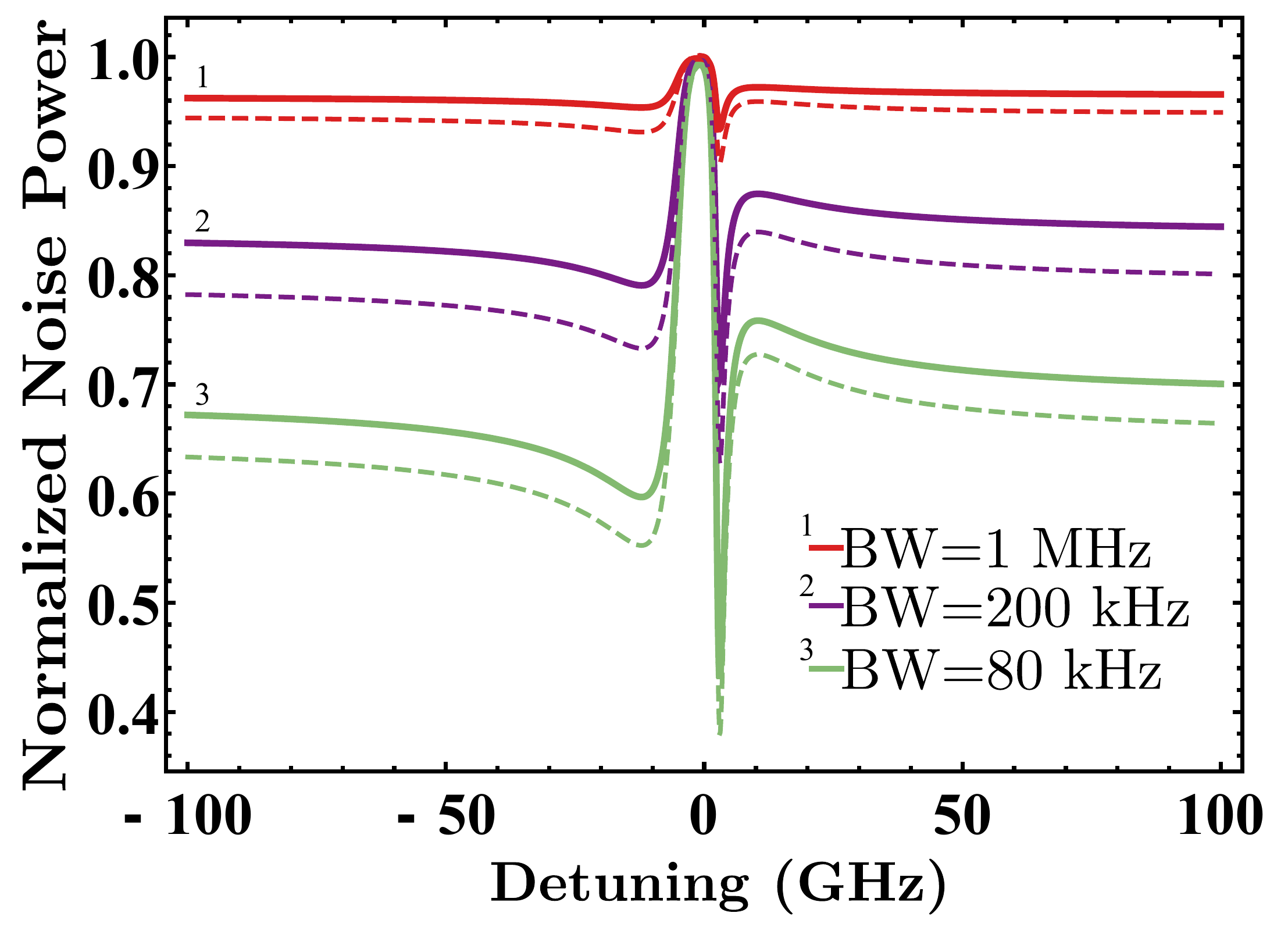} 
\caption{Integrated spin-noise power as a function of the detuning. The spectrum is integrated up to three different bandwidths, $1$ \si{\mega\hertz} (no.1-red), $200$ \si{\kilo\hertz} (no.2-purple) and $80$ \si{\kilo\hertz} (no3.-green). The bandwidth is related to the sampling time through $\mathrm{BW}=1/(2 \Delta t)$. Solid lines correspond to direct integration of the power spectrum up to the bandwidth frequency, whereas the dashed lines correspond to the noise given by Eq.\eqref{eq:NoisePowerSpecBandwidth}. The spectrum corresponds to a $^{87}$Rb vapor at a temperature of $T=200$ $^{\circ}$C and a magnetic field of $B=10$ mG. The optical linewidth is $\Delta \nu=10$ \si{\giga\hertz}. The response is normalized to the total atomic noise $g_a^2 \var(F^a)  + g_b^2  \var(F^b)$ obtained when integrating the spectrum in the whole frequency range.}
\label{fig:noise_det}
\end{figure}

To further illustrate these features, in Fig. \ref{fig:noise_det} we plot the spin-noise power as a function of the detuning for three different bandwidths. The power is obtained in two different ways: either by integrating the spectrum up to the frequency defined by BW or by directly applying Eq.\eqref{eq:NoisePowerSpecBandwidth}. In both cases, the power has been normalized against the total power in the whole frequency range, given by Eq.\ref{Eq:SN-power}. It is apparent that both methods produce similar results.
Interestingly, for a $^{87}$Rb vapor at realistic conditions, e.g., a cell temperature of $T=200$ $^{\circ}$C, a magnetic field of $B=10$ mG, a pressure broadening of $\Delta \nu=10$ \si{\giga\hertz}, and a measurement bandwidth of $200$ \si{\kilo\hertz}, we obtain a spin-noise power at the far-wings of the optical transition that is approximately $20$ \% smaller than the noise of the thermal state. Apparently, for even smaller acquisition bandwidths, it can exceed $40$ \%. In the limit where the bandwidth extends to infinity, we obtain $g_a^2 \var(F^a)  + g_b^2  \var(F^b)$. The same results are also obtained when the noise-power is extracted from the time domain signal since there is a direct relation between Eqs. \eqref{eq:CorrectedVarTimeWindow} and \eqref{eq:NoisePowerSpecBandwidth}.

\begin{figure*}[htp]
\centering
\includegraphics[width=15cm]{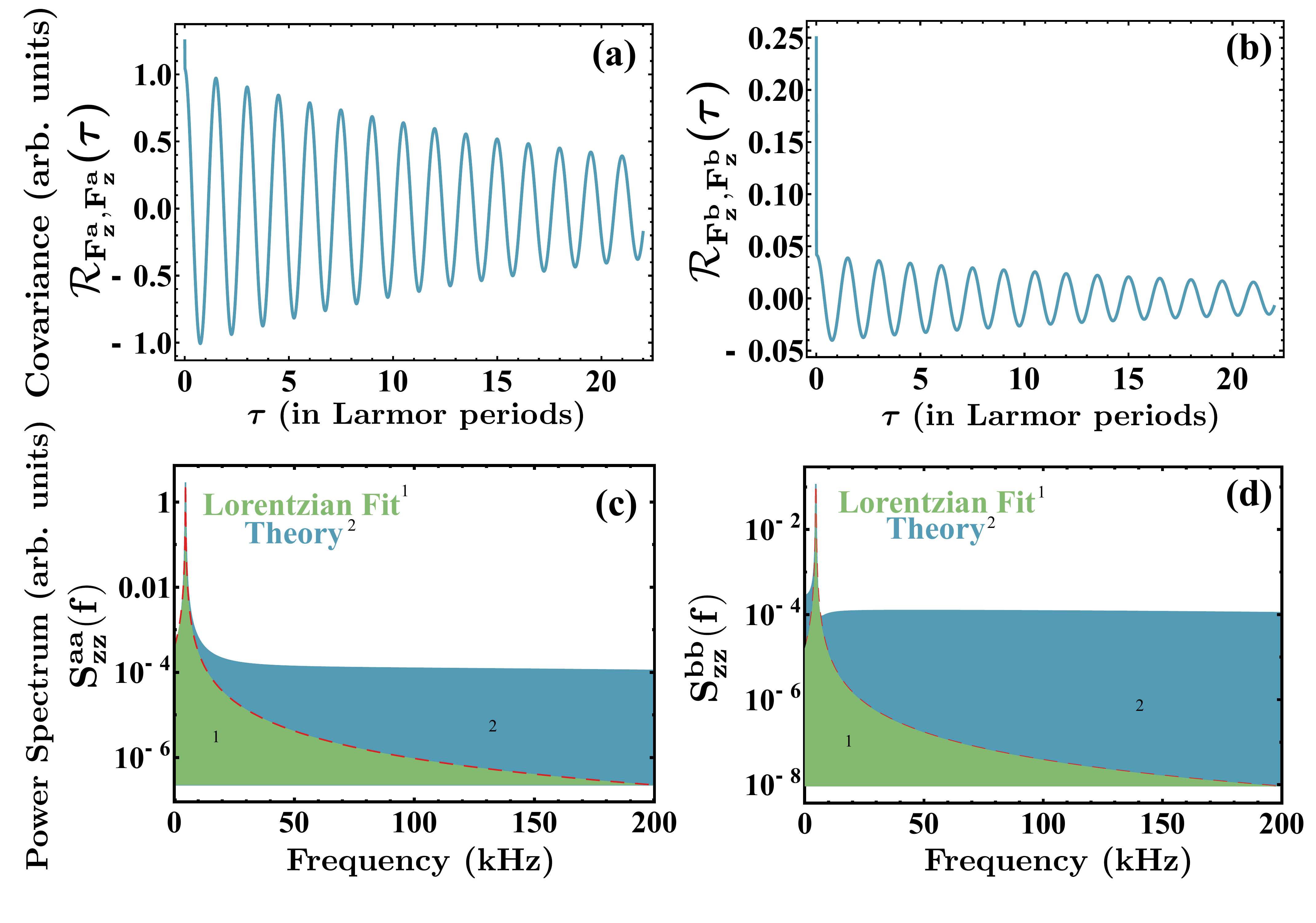} 
\caption{$\tau$-evolution of a) $\mathcal{R}_{\hat{F}_z^{a},\hat{F}_z^{a}} (\tau)$ and b) $\mathcal{R}_{\hat{F}_z^{b},\hat{F}_z^{b}} (\tau)$ with the corresponding power spectra presented in c) and d). The green filled areas (no.1) represent a Lorentzian fit to the power spectrum of the form $\gamma/[(\nu-\nu_0)^2 +\gamma^2]$. The spectrum corresponds to a $^{87}$Rb vapor at a magnetic field of $B=10$ mG and a spin-exchange rate 100 times larger than the Larmor frequency.}
\label{fig:fit spectrum}
\end{figure*}

Lastly, we emphasize that fitting the spectrum with a single Lorentzian, leads to a considerable underestimation of the spin-noise power, given by the integrated area under the Lorentzian curve. This is illustrated in Fig. \ref{fig:fit spectrum} where both of the spectra in the two hyperfine states are fitted by simple Lorentzian functions, depicted by the green filled areas (no.1). We find that in the upper manifold the fit captures $83$ \% of the total power, whereas in the lower manifold only $16$ \% is captured. This is expected since the first is dominated by the slow and the latter by the fast relaxation, as demonstrated in Figs. \ref{fig:fit spectrum}a) and b). In the SERF regime, the coupling between the two hyperfine multiplets is roughly characterized by $\Gamma_{-}=(\kappa_{aa}+\kappa_{bb})R_{\mathrm{se}}$ \cite{PhysRevA.103.043116}, therefore it takes at least $1/\Gamma_{-}$ amount of time for the two manifolds to be correlated. This is reflected in the rapid jumps of the correlation functions close to $\tau=0$. After that time interval, because of the strong interlock of the two states, the resulting correlations have long coherence times with amplitudes given in \autoref{fig:cov_coefficients}.

\subsection{Observations}
\label{sec:Obervations}

The new features are the result of strong correlations between the two hyperfine manifolds in the SERF regime. These correlations result from the rapid transfer of angular momentum between the two hyperfine populations, and they cause the two states to precess in an interlocked way.  

The effects are most visible when the probe frequency is tuned to the wings of the lower hyperfine level. Stronger effects are expected for low buffer gas pressures, for which the optical transitions from the ground-state hyperfine levels are better resolved.

For the shake of simplicity, we have not included spin-destruction collisions, power broadening, or diffusion in the spin dynamics model. The first two of these can be included quite simply in the Bloch equations.  Diffusion can also be included, using a mode expansion \cite{PhysRevA.102.012822}, at the cost of increasing the dimensionality of the model. 

As just noted, the most dramatic effects occur at low buffer gas pressure, and thus in conditions for which spin destruction collisions play a minor role.  Similarly, diffusion and power broadening can be made negligible by working with a large probe beam size \cite{PhysRevA.96.062702}.  For these reasons, we believe the effects predicted here should be observable in appropriate conditions, and that other conditions can also be modeled through straightforward adaptation of the theory given here. 

To the best of our knowledge, there is no simple model (simpler than the coupled hyperfine Bloch equations we utilize here) that can accurately describe the atomic noise properties in the SERF regime or in the transition between SERF and SE regimes.   
Here and in the abstract, by ``simple model'' we refer to a Bloch equation describing a single spin (or collective spin) observable, e.g.,
\begin{equation}
\frac{d \mathbf{S}}{dt} = \frac{1}{q(P)} (\gamma_e \mathbf{S} \times \mathbf{B}) -\Gamma_{\rm{rel}} \mathbf{S} + \sqrt{2 \Gamma_{\rm{rel}} \sigma} d\mathbf{W}     
\end{equation}
where $\mathbf{S}=(S_x,S_y,S_z)$ is the electron spin, $\gamma_e=g_s \mu_{_B}/ \hbar$ the gyromagnetic ratio of the electron, $q(P)$ is the nuclear slowing-down factor that depends on the spin-polarization, $\Gamma_{\rm{rel}}$ the spin relaxation rate, $\sigma$ is the equilibrium variance (spin-projection noise) and $d\mathbf{W}=(dW_x,dW_y,dW_z)$ is a vector with independent temporal Wiener increments. 

Such models have been frequently used in spin-noise spectroscopy with unpolarized vapors. Assuming that the spin relaxation is only related to spin-exchange collisions, there is a regime where the simple Bloch model is approximately valid. This regime is at large magnetic fields, such that the Larmor frequency is much larger than the spin-exchange rate $R_{\mathrm{se}}$.  In this scenario, the relaxation rates $\Gamma_{\pm}$ that appear in Eq.\eqref{eq:cov-tau}, are nearly equal, each being roughly equal to $R_{\mathrm{se}}$ and therefore, the power spectrum can be well approximated by a single Lorentzian with an effective relaxation rate and with the atomic spin-noise power concentrated at the spin-precession frequency. This conclusion holds for all detunings and buffer gas pressures. Additionally, at high magnetic fields, spin-correlations between the two hyperfine levels are negligible and the dynamics of the two manifolds can be considered independently. The above assumptions motivate the use of a simple Bloch model in this large field regime.

The simple model fails in the SERF regime, because of the fast relaxation rate $\Gamma_{+}$ that causes the atomic noise to be distributed to a broad range of frequencies, and also because ``broadband'' atomic noise is transferred to the cross-correlation spectrum between the two hyperfine manifolds. In that case, as we demonstrate in \autoref{fig:fit spectrum}, a simple model with a simple Lorentzian spectrum does not accurately describe the spin-noise spectrum and the spin-noise power. Because of the broadband atomic noise components, the measurement bandwidth becomes important when calculating the integrated noise power in the SERF regime. In this regime, a more complete description, like the one we have developed here, is required.

\section{Conclusions}
\label{sec:Conclusions}

We have derived analytic expressions for the spin-noise correlations and spin-noise spectra of hot alkali-metal vapors, taking into account hyperfine, Zeeman, and spin-exchange interactions, and also accounting for the effects of probe detuning relative to the different atomic  hyperfine transitions. The results are applicable both in and out of the SERF regime, and show several interesting spectral behaviors not seen in phenomenological models. 

The observation that spin-noise can be reduced at the Larmor frequency by proper choice of detuning is particularly intriguing for magnetic sensing and extensions such as magnetic gradiometry and comagnetometry.

\appendix

\section{Spin-dynamics}
\label{sec:spin-dyn}
The relaxation of the angular momentum in the ground-state of the alkali-metal atoms due to spin-exchange collisions in a constant magnetic field is described by the non-linear differential equation for the density matrix,  Eq.\eqref{eq:density-matrix}. Defining the single-atom projector operators in the upper and lower hyperfine manifolds as $\hat{P}_{\alpha}= \sum_{m} \ket{\alpha m} \bra{\alpha m}$ with $\alpha \in \{ a,b \}$, the density matrix can be decomposed as $\rho= \mathbb{1} \rho \mathbb{1}= \sum_{\alpha}\hat{P}_{\alpha} \rho \hat{P}_{\alpha} \approx \hat{P}_{a} \rho \hat{P}_{a}+ \hat{P}_{b} \rho \hat{P}_{b} \equiv \rho_a +\rho_b$. In arriving at the last step, we have ignored hyperfine coherences since they exceed by far the bandwidth of interest in this work.

In the low-polarization limit, the equation can be considerably simplified, allowing one to write the evolution of  $\langle\mathbf{\hat{F}^a}\rangle=\text{Tr}[\tilde{\rho} \mathbf{\hat{F}^a}]$ and $\langle \mathbf{\hat{F}^b}\rangle=\text{Tr}[\tilde{\rho} \mathbf{\hat{F}^b}]$ into a  system of first-order coupled differential equations as demonstrated in Eqs.\eqref{eq:DiffEqA} and \eqref{eq:DiffEqB}. This is realized by expanding the single atom density matrix $\rho$ in the spherical tensor operator basis 
 $\rho = \sum_{_{LMff'}} \rho_{_{LMff'}} {\rm T}^L_M(ff')=\sum_{\Lambda \mu l m} \rho_{_{\Lambda \mu l m}} {\rm T}^{\Lambda}_{\mu} (II) \otimes {\rm T}^l_m (ss)$ either in the coupled or in the uncoupled basis, spanned by the angular momentum states $\{ \ket{fm} \}$ and $\{ \ket{sm_s} \otimes \ket{Im_{_I}} \}$, respectively.
Here ${\rm T}_M^L(f f')$ are the irreducible spherical tensor operators defined as \cite{PhysRev.163.12}
\begin{equation}
\begin{split}
{\rm T}_M^L(f f')= &\sum_m (-1)^{m-M-f'} \ket{fm} \bra{f' m-M} \times\\
&\times C(ff' L;m,M-m),
\end{split}
\end{equation}
where $f,f' \in \{a,b\}$ are the two ground-state hyperfine states and $C(ff' L;m,M-m)$ denotes the Clebsch-Gordan coefficient. For the coefficients $\kappa_{\alpha \beta}$ with $\alpha,\beta \in \{ a,b\}$ appearing in Eqs.\eqref{eq:DiffEqA} and \eqref{eq:DiffEqB}, one obtains
\begin{equation}
\begin{split}
\kappa_{aa} & = 1-X_1(aa)^2 - Y_1(aa)^2,\\
\kappa_{ab}&=r[X_1(aa)X_1(bb)+Y_1(aa)Y_1(bb)],\\
\kappa_{ba}&=\frac{1}{r}[]X_1(aa)X_1(bb)+Y_1(aa)Y_1(bb)],\\
\kappa_{bb}&=1-X_1(bb)^2-Y_1(bb)^2.
\end{split}
\end{equation}
where  $ r=\sqrt{a(a+1)(2a+1)/[b(b+1)(2b+1)]}$. The coefficients $X_1(aa), X_1(bb),Y_1(aa)$ and $Y_1(bb)$ are frequently encountered in the theory of SE in alkali-metal atoms when moving between the uncoupled and the coupled spherical tensor operator bases, and they are given by the expressions \cite{HapperRev} 
\begin{equation}
X_L(ff')={( - 1)^{ - (f' + s + L + I)}} \Big ( \frac{[f][f']}{[s]} \Big )^{1/2} \left\{ {\begin{array}{*{20}{c}}
  f'&s&I \\ 
  I&L&f 
\end{array}} \right\},
\end{equation}
\begin{equation}
Y_L(ff')={( - 1)^{ - (I + f + s + L)}} \Big ( \frac{[f][f']}{[I]} \Big )^{1/2} \left\{ {\begin{array}{*{20}{c}}
  I&f&s \\ 
  L&s&f' 
\end{array}} \right\}.
\end{equation}
To avoid redundancy here we have reproduced the key results of \cite{PhysRevA.103.043116} that will be utilized throughout the paper. The derivation of these formulas is stemming directly from Happer-Tam's seminal paper \cite{PhysRevA.16.1877} and has been recently analyzed in \cite{katz2015coherent,PhysRevA.103.043116}. In terms of the nuclear spin, the coefficients $\kappa_{\alpha \beta}$ take the form
\begin{equation}
\begin{split}
\kappa_{aa} & =\frac{2}{3}\frac{I(2I-1)}{(2I+1)^2},\\
\kappa_{bb}&= \frac{2}{3}\frac{(2I+3)(I+1)}{(2I+1)^2},\\
\kappa_{ab}&=r\frac{2}{3}\frac{[I(I+1)(2I-1)(2I+3)]^{1/2}}{(2I+1)^2},\\
\kappa_{ba}&=\frac{1}{r}\frac{2}{3}\frac{[I(I+1)(2I-1)(2I+3)]^{1/2}}{(2I+1)^2}.\\
\end{split}
\end{equation}
We note that $\kappa_{ab}=r\sqrt{\kappa_{aa}\kappa_{bb}}$, $\kappa_{ba}=\sqrt{\kappa_{aa}\kappa_{bb}}/r$ and $r=\sqrt{\kappa_{bb}/\kappa_{aa}}$, therefore we conclude that $\kappa_{ab}=\kappa_{bb}$ and $\kappa_{ba}=\kappa_{aa}$.

\section{Eigenvalues}
\label{sec:Eigenspectrum}
Out of the six eigenvalues of $A$, two are real ($\lambda_5=0,   \lambda_6=-(\kappa_{aa}+\kappa_{bb}) R_{\mathrm{se}}$) and associated with the longitudinal angular momentum components, whereas the remaining four are complex and related to the transverse components. The latter can be compactly written in a simple formula. Defining $\kappa_{+} \equiv \kappa_{aa}+\kappa_{bb}$ and $\kappa_{-} \equiv \kappa_{aa}-\kappa_{bb}$ we obtain
\begin{equation}
\lambda =- \Big[\frac{1}{2}\kappa_{+}R_{\mathrm{se}} \pm \frac{1}{2}\sqrt{(\kappa_{+} R_{\mathrm{se}})^2 \pm 4i\kappa_{-} R_{\mathrm{se}}\omega_0-4\omega_0^2} \Big] . 
\end{equation}
The real part gives the relaxation rate of the transverse spin components, whilst the imaginary part the precession frequency in the magnetic field.
The quantity under the square root is a complex number of the form $z_r =x_r+iy_r$, with $x_r=(\kappa_{+} R_{\mathrm{se}})^2-4\omega_0^2$ and $y_r=\pm4\kappa_{-} R_{\mathrm{se}}\omega_0$. The square root function is also a complex number that can be expanded in terms of the real and imaginary parts as
\begin{equation}
\begin{split}
&X+iY =\sqrt{x_r+iy_r}\\
&= \sqrt{\sqrt{x_r^2+y_r^2}+x_r}+i \text{Sign}[y_r]\sqrt{\sqrt{x_r^2+y_r^2}-x_r}.
\end{split}
\end{equation}
In this way, the real and imaginary parts of the eigenvalues are separated and consequently one obtains $\lambda_1 =- (\Gamma_{-} + i \Omega_{-})$, $\lambda_2 = -(\Gamma_{+} + i \Omega_{+})$, $\lambda_3 = \lambda_1^{*} =-(\Gamma_{-} - i \Omega_{-})$ and $\lambda_4 = \lambda_2^{*}= -(\Gamma_{+} - i \Omega_{+})$. Here we have defined $\Gamma_{\pm}\equiv (\kappa_{+}R_{\mathrm{se}} \pm X)/2$ and $\Omega_{\pm}\equiv \pm Y/2$ being the relaxation rates and precession frequencies in the low-polarization limit when both the spin-exchange rate $R_{\mathrm{se}}$ and the Larmor frequency $\omega_0$ are much slower than the hyperfine rate. 

Furthermore, it should be noted that the complex eigenvalues can be expressed in terms of the nuclear spin multiplicity $[I] \equiv 2I+1$ as
\begin{equation}
\begin{split}
\lambda &= -\Big[
\frac{[I]^2+2}{3[I]^2}R_{\mathrm{se}} \pm\\
&\pm\sqrt{\Big(\frac{([I]^2+2)}{3[I]^2}\Big)^2R_{\mathrm{se}}^2 \pm\frac{2iR_{\mathrm{se}}\omega_0}{[I]}-\omega_0^2 } \Big],
\end{split}
\end{equation}
in agreement with Eq.(99) of \cite{PhysRevA.16.1877}. Analytical expressions for $\Gamma_{\pm}$ and $\Omega_{\pm}$ have been also presented in the same paper both in the rapid ($ \omega_{0} \ll R_{\mathrm{se}}$) and in the slow ($ \omega_{0} \gg R_{\mathrm{se}}$) SE regimes. 

The relaxation rate and the precession frequency of the magnetic resonance spectrum depend also on the spin-polarization. At high atomic polarization $P \approx 1$, due to optical pumping, most of the atoms have populated a stretched state with $m=\pm f$ and SE collisions are insignificant there \cite{SavukovRomalis}. This is the case since all the atoms have the same spin projection and the exchange interaction is symmetric in that case.  Consequently, the gyromagnetic ratio does not depend on the ratio between $R_{\mathrm{se}}$ and $\omega_0$ and therefore the precession frequency is defined considering only the Hamiltonian dynamics i.e., $\gamma=\gamma_0=g_s \mu_{_B}/ \hbar (2I+1)$. However, for intermediate or low spin-polarization, the gyromagnetic ratio becomes generally smaller as the SERF regime is approached. The value of $\gamma$ in SERF scales with polarization in a polynomial fashion and the higher the spin-polarization the closer to $\gamma_0$ the gyromagnetic ratio is. For instance, the gyromagnetic ratio  of a $^{87}$Rb vapor scales with polarization as $\gamma \approx \gamma_e/ q(P)$ where $q(P)=(6+2P^2)/(1+P^2)$ is the so-called nuclear slowing down factor \cite{Appelt}. In the lowest polarization limit where the magnetic sublevels are equally populated, the slowing down factor becomes equal to the ratio of the electron gyromagnetic ratio to the atomic gyromagnetic ratio $[s(s+1) + I(I+1)]/s(s+1)$.

\section{Measurement bandwidth}
\label{sec:MB}
Assuming that we sample the stochastic atomic signal with a time step $\Delta t$, then the bandwidth of the sampling process is $\mathrm{BW}=1/(2\Delta t)$. In a time-interval $[t,t+ \Delta t]$ we measure the time-averaged signal
\begin{equation}
\bar{\hat{F}}_i^{\alpha} (t) = \frac{1}{\Delta t} \int_{t}^{t+\Delta t} \hat{F}_i^{\alpha} (t') dt'  .
\end{equation}
Then, the average covariance of the acquired signal in this time interval is given by 
\begin{widetext}
\begin{align}
\bar{\mathcal{R}}_{\hat{F}_i^{\alpha},\hat{F}_j^{\beta}} &=\frac{1}{\Delta t^2} \mathrm{Re}\Big[\int_{t}^{t+\Delta t} \int_{t}^{t+\Delta t} \langle \hat{F}_i^{\alpha} (t') \hat{F}_j^{\beta} (t'') \rangle dt' dt'' \Big] =\frac{1}{\Delta t^2} \mathrm{Re}\Big[  \int_{t}^{t+\Delta t} \int_{t}^{t+\Delta t} \langle \hat{F}_i^{\alpha} (t'-t'') \hat{F}_j^{\beta} (0) \rangle dt'dt''\Big] \\
& =\frac{1}{\Delta t^2} \mathrm{Re}\Big[ \int_{t}^{t+\Delta t} dt' \left[ \int_{t}^{t'} dt'' \langle \hat{F}_i^{\alpha} (t'-t'') \hat{F}_j^{\beta} (0)\rangle+\int_{t'}^{t+\Delta t} dt'' \langle \hat{F}_i^{\alpha} (t'-t'') \hat{F}_j^{\beta} (0)\rangle \right]\Big].
\end{align}
\end{widetext}
The above integral can be expressed in terms of the correlation function $\mathcal{R}_{F_i^{\alpha},F_j^{\beta}}(\tau)$ through changing the variables of integration from $(t',t'')$ to $(\tau=t'-t'',s=t'') $. The change of variables leads to a modification in the integral limits and consequently we find

\begin{widetext}
\begin{align}
\bar{\mathcal{R}}_{\hat{F}_i^{\alpha},\hat{F}_j^{\beta}} &= \frac{1}{\Delta t^2} \mathrm{Re}\Big[  \int_{t}^{t+\Delta t} \int_{t}^{t+\Delta t} \langle \hat{F}_i^{\alpha} (t'-t'') \hat{F}_j^{\beta} (0) \rangle dt'dt''\Big]= \frac{1}{\Delta t^2} \mathrm{Re}\Big[  \int \int \langle \hat{F}_i^{\alpha} (\tau) \hat{F}_j^{\beta} (0) \rangle d\tau ds \Big]\\
&=\frac{1}{\Delta t^2} \mathrm{Re} \Big \{ \int_{-\Delta t}^{0} d\tau \int_{t-\tau}^{t+\Delta t} ds \langle \hat{F}_i^{\alpha} (\tau) \hat{F}_j^{\beta} (0) \rangle+ \int_{0}^{\Delta t} d\tau \int_{t}^{t+\Delta t-\tau} ds \langle \hat{F}_i^{\alpha} (\tau) \hat{F}_j^{\beta} (0) \rangle \Big \} \\
&= \frac{1}{\Delta t^2} \mathrm{Re} \Big \{ \int_{-\Delta t}^{0} d\tau \mathcal{R}_{\hat{F}_i^{\alpha},\hat{F}_j^{\beta}}(\tau) \left[ \Delta t +\tau \right] + \int_{0}^{\Delta t} d\tau \mathcal{R}_{\hat{F}_i^{\alpha},\hat{F}_j^{\beta}}(\tau) \left[ \Delta t -\tau \right] \Big \} .
\end{align}
\end{widetext}

\section{$\tau$-dependence of the steady-state covariance}
\label{sec:time evolution}
As described in \autoref{sec:Unequal-time correlations}, analytical expressions for the unequal-time correlations can be obtained by applying the exponential decomposition $e^{A \tau}= V e^{\Lambda \tau} V^{-1}$. By explicitly performing the matrix multiplications, we find that each covariance component can be written as a weighted sum of all the exponentiated eigenvalues of $A$ as
\begin{equation}
\mathcal{R}_{\hat{F}_i^{\alpha},\hat{F}_j^{\beta}}(\tau) =
\begin{cases}
\sum_{n} c_n(F^{\alpha}_{i},F^{\beta}_{j}) e^{\lambda_n \tau} & ,\tau > 0\\
\\
\sum_{n} q_n(F^{\alpha}_{i},F^{\beta}_{j}) e^{-\lambda_n \tau} & ,\tau < 0
 \end{cases}
\label{eq:cov_cases}
\end{equation}
The coefficients $c_n(F^{\alpha}_{i},F^{\beta}_{j})$ and $q_n(F^{\alpha}_{i},F^{\beta}_{j})$ are the result of the previously discussed matrix multiplications. Evidently, in the dynamics of the longitudinal components, the coefficients weighting the eigenvalues associated with the transverse spin components are zero and vice versa. This observation is complemented by the fact that there are zero correlations between the longitudinal and transverse spin components. For example, in Eqs.\eqref{eq:longitudinal1} -- \eqref{eq:longitudinal4} it is apparent that only $\lambda_5=0$ and $\lambda_6=- (\kappa_{aa}+\kappa_{bb})R_{\mathrm{se}}$ contribute in the $\tau$-dependence. Moreover, taking into account emerging conjugation relations e.g., $c_3(F^{\alpha}_{i},F^{\beta}_{j})= c_1^{*}(F^{\alpha}_{i},F^{\beta}_{j})$ and $c_4(F^{\alpha}_{i},F^{\beta}_{j})= c_2^{*}(F^{\alpha}_{i},F^{\beta}_{j})$ associated with the conjugation relation of the eigenvalues $\lambda_3= \lambda_1^{*}$ and $\lambda_4 = \lambda_2^{*}$, we arrive at  Eq.\eqref{eq:cov-tau}, where only eigenvalues associated with the transverse spin components have a non-zero contribution.

We recall from Appendix \ref{sec:Eigenspectrum} that $\kappa_{-}\equiv (\kappa_{aa}-\kappa_{bb})R_{\mathrm{se}}$, and $X$ and $Y$ are the real and imaginary parts of the square root entering in the expressions for the eigenvalues. Then, the coefficients of the $\tau$-evolution of the steady-state covariance matrix elements entering in the expression for the Faraday signal are given by
\begin{equation}
c_1(F^{a}_{z},F^{a}_{z}) =\frac{1}{4} \var(F^{a}) \Big[1-\frac{\kappa_{-}R_{\mathrm{se}}+2i\omega_0}{X+iY} \Big],
\end{equation}

\begin{equation}
c_2(F^{a}_{z},F^{a}_{z}) =\frac{1}{4} \var(F^{a}) \Big[1+\frac{\kappa_{-}R_{\mathrm{se}}+2i\omega_0}{X+iY} \Big],
\end{equation}

\begin{equation}
c_1(F^{a}_{z},F^{b}_{z}) =\frac{1}{2} \var(F^{b}) \frac{\kappa_{bb}R_{\mathrm{se}}}{X+iY},
\end{equation}

\begin{equation}
c_2(F^{a}_{z},F^{b}_{z}) =-\frac{1}{2} \var(F^{b}) \frac{\kappa_{bb}R_{\mathrm{se}}}{X+iY},
\end{equation}

\begin{equation}
c_1(F^{b}_{z},F^{a}_{z}) =\frac{1}{2} \var(F^{a}) \frac{\kappa_{aa}R_{\mathrm{se}}}{X+iY},
\end{equation}

\begin{equation}
c_2(F^{b}_{z},F^{a}_{z})=-\frac{1}{2} \var(F^{a}) \frac{\kappa_{aa}R_{\mathrm{se}}}{X+iY},
\end{equation}

\begin{equation}
c_1(F^{b}_{z},F^{b}_{z}) =\frac{1}{4} \var(F^{b}) \Big[1+\frac{\kappa_{-}R_{\mathrm{se}}+2i\omega_0}{X+iY} \Big],
\end{equation}

\begin{equation}
c_2(F^{b}_{z},F^{b}_{z}) =\frac{1}{4} \var(F^{b}) \Big[1-\frac{\kappa_{-}R_{\mathrm{se}}+2i\omega_0}{X+iY} \Big].
\end{equation}
We note that $\kappa_{aa} \var{(F^a)}=\kappa_{bb} \var{(F^b)}$ so that the off-diagonal coefficients are equal. Similar expressions can be obtained for the rest of the covariance matrix elements not discussed in this paper, but also for the conjugate coefficients $q_n(F^{\alpha}_{i},F^{\beta}_{j})$ obtained by decomposing the second branch of Eq.\eqref{eq:cov}. At zero magnetic field we find $X\rightarrow \kappa_{+}R_{\mathrm{se}}$ and $Y \rightarrow 0$, therefore the coefficients are significantly simplified.

As a side note, we point out that we can  express the power spectrum in terms of $c_n(F^{\alpha}_{i},F^{\beta}_{j})$ and $q_n(F^{\alpha}_{i},F^{\beta}_{j})$. Plugging Eq.\eqref{eq:cov_cases} into Eq.\eqref{eq:power spec} and performing the integration over $\tau$, we find that the matrix elements of the power spectrum, regarding the atomic signal result in 
\begin{equation}
S_{i,j}^{\alpha,\beta}(\omega)= \sum_{n} \frac{q_n(F^{\alpha}_{i},F^{\beta}_{j})}{-\lambda_n-i\omega}   +\frac{c_n(F^{\alpha}_{i},F^{\beta}_{j})}{-\lambda_n+i\omega}.  \label{eq:PowerSpec}
\end{equation}

Finally, we demonstrate a different representation of the correlation function which is frequently encountered in SNS, especially when simplified Bloch models are utilized. Eventually, we can bring Eq.\eqref{eq:cov-tau} in a more familiar form by noticing that $\pm [a \cos(x) + b \sin(x) ]= \pm h \cos(x - \theta)$ where $h=\sqrt{a^2 +b^2}$ and $\tan(\theta)=b/a$. Applying the preceding identity, we obtain for $\tau>0$,
\begin{equation}
\mathcal{R}_{F_i^{\alpha},F_j^{\beta}}(\tau) =
\sum_{q=\pm} \mathrm{Sign}(\mathrm{Re}[c_{q}]) |c_{q}| e^{-\Gamma_{q} \tau}  \cos{(\Omega_{q}\tau - \theta_{q})}, \label{eq:cov-tau2}
\end{equation}
with $\theta_{\pm} = \tan^{-1}( \mathrm{Im}[c_{\pm}]/ \mathrm{Re}[c_{\pm}])$.  Here we have made the following replacements to the coefficients $2 c_{1}\rightarrow c_{-}$ and $2 c_2 \rightarrow c_{+}$. 

The amplitudes $|c_{\pm}|$ and the phases $\theta_{\pm}$ depend on the ratio $R_{\mathrm{se}}/\omega_0$. Both amplitudes $|c_{\pm}|$ add up to produce the total amplitude of the oscillating covariance. In addition, the transition from the slow SE regime to SERF is marked by a change of $2 \pi$ in the phase of the auto-correlations $ \mathcal{R}_{F_z^{a},F_z^{a}}(\tau)$ and $ \mathcal{R}_{F_z^{b},F_z^{b}}(\tau)$ and a change of $\pi$ in the cross-correlations $\mathcal{R}_{F_z^{a},F_z^{b}}(\tau) $ and $\mathcal{R}_{F_z^{b},F_z^{a}}(\tau)$. Finally, we note that the time evolution of the covariance components is identical for $\tau<0$, however the amplitudes $|c_q|$ and the phases $\theta_q$ will generally differ from those at $\tau>0$.

\section*{Acknowledgement}
We thank Charikleia Troullinou for fruitful discussions. KM, VGL and MWM acknowledge funding from H2020 Future and Emerging Technologies Quantum Technologies Flagship projects MACQSIMAL (Grant Agreement No. 820393) 
H2020 Marie Sk{\l}odowska-Curie Actions project ITN ZULF-NMR  (Grant Agreement No. 766402);  
Spanish Ministry of Science Plan de Recuperaci\'{o}n, Transformac\'{\i}on y Resiliencia - financed by the European Union - NextGenerationEU and  ``Severo Ochoa'' Center of Excellence CEX2019-000910-S, and projects OCARINA (PGC2018-097056-B-I00) and  SAPONARIA (PID2021-123813NB-I00),  funded by MCIN/ AEI /10.13039/501100011033/ FEDER ``A way to make Europe'' ; Generalitat de Catalunya through the CERCA program; 
Ag\`{e}ncia de Gesti\'{o} d'Ajuts Universitaris i de Recerca Grant No. 2017-SGR-1354;  Secretaria d'Universitats i Recerca del Departament d'Empresa i Coneixement de la Generalitat de Catalunya, co-funded by the European Union Regional Development Fund within the ERDF Operational Program of Catalunya (project QuantumCat, ref. 001-P-001644); Fundaci\'{o} Privada Cellex; Fundaci\'{o} Mir-Puig; Marie Sk{\l}odowska-Curie Actions project PROBIST (Grant Agreement No. 754510). GV acknowledges funding from EU QuantERA Project PACE-IN (GSRT Grant No. T11EPA4-00015) and from the Hellenic Foundation for Research and Innovation (HFRI) and the General Secretariat for Research and Technology (GSRT), under Grant Agreement No. [00768]. JK thanks the support from National Natural Science Foundation of China (NSFC) (Grant No. 12005049 and No. 11935012). IKK acknowledges the co-financing of this research by the European Union and Greek national funds through the Operational Program Crete 2020-2024, under the call ``Partnerships of Companies with Institutions for Research and Transfer of Knowledge in the Thematic Priorities of RIS3Crete,'' with project title ``Analyzing urban dynamics through monitoring the city magnetic environment'' (Project KPHP1—No. 0029067). 

\bibliography{references}

\begin{thebibliography}{60}%
\makeatletter
\providecommand \@ifxundefined [1]{%
 \@ifx{#1\undefined}
}%
\providecommand \@ifnum [1]{%
 \ifnum #1\expandafter \@firstoftwo
 \else \expandafter \@secondoftwo
 \fi
}%
\providecommand \@ifx [1]{%
 \ifx #1\expandafter \@firstoftwo
 \else \expandafter \@secondoftwo
 \fi
}%
\providecommand \natexlab [1]{#1}%
\providecommand \enquote  [1]{``#1''}%
\providecommand \bibnamefont  [1]{#1}%
\providecommand \bibfnamefont [1]{#1}%
\providecommand \citenamefont [1]{#1}%
\providecommand \href@noop [0]{\@secondoftwo}%
\providecommand \href [0]{\begingroup \@sanitize@url \@href}%
\providecommand \@href[1]{\@@startlink{#1}\@@href}%
\providecommand \@@href[1]{\endgroup#1\@@endlink}%
\providecommand \@sanitize@url [0]{\catcode `\\12\catcode `\$12\catcode
  `\&12\catcode `\#12\catcode `\^12\catcode `\_12\catcode `\%12\relax}%
\providecommand \@@startlink[1]{}%
\providecommand \@@endlink[0]{}%
\providecommand \url  [0]{\begingroup\@sanitize@url \@url }%
\providecommand \@url [1]{\endgroup\@href {#1}{\urlprefix }}%
\providecommand \urlprefix  [0]{URL }%
\providecommand \Eprint [0]{\href }%
\providecommand \doibase [0]{https://doi.org/}%
\providecommand \selectlanguage [0]{\@gobble}%
\providecommand \bibinfo  [0]{\@secondoftwo}%
\providecommand \bibfield  [0]{\@secondoftwo}%
\providecommand \translation [1]{[#1]}%
\providecommand \BibitemOpen [0]{}%
\providecommand \bibitemStop [0]{}%
\providecommand \bibitemNoStop [0]{.\EOS\space}%
\providecommand \EOS [0]{\spacefactor3000\relax}%
\providecommand \BibitemShut  [1]{\csname bibitem#1\endcsname}%
\let\auto@bib@innerbib\@empty
\bibitem [{\citenamefont {Budker}\ and\ \citenamefont
  {Romalis}(2007)}]{budker2007optical}%
  \BibitemOpen
  \bibfield  {author} {\bibinfo {author} {\bibfnamefont {D.}~\bibnamefont
  {Budker}}\ and\ \bibinfo {author} {\bibfnamefont {M.}~\bibnamefont
  {Romalis}},\ }\bibfield  {title} {\bibinfo {title} {Optical magnetometry},\
  }\href {https://doi.org/10.1038/nphys566} {\bibfield  {journal} {\bibinfo
  {journal} {Nature Physics}\ }\textbf {\bibinfo {volume} {\textbf{3}}},\
  \bibinfo {pages} {227} (\bibinfo {year} {2007})}\BibitemShut {NoStop}%
\bibitem [{\citenamefont {Limes}\ \emph {et~al.}(2020)\citenamefont {Limes},
  \citenamefont {Foley}, \citenamefont {Kornack}, \citenamefont {Caliga},
  \citenamefont {McBride}, \citenamefont {Braun}, \citenamefont {Lee},
  \citenamefont {Lucivero},\ and\ \citenamefont {Romalis}}]{LimesPRAppl2020}%
  \BibitemOpen
  \bibfield  {author} {\bibinfo {author} {\bibfnamefont {M.}~\bibnamefont
  {Limes}}, \bibinfo {author} {\bibfnamefont {E.}~\bibnamefont {Foley}},
  \bibinfo {author} {\bibfnamefont {T.}~\bibnamefont {Kornack}}, \bibinfo
  {author} {\bibfnamefont {S.}~\bibnamefont {Caliga}}, \bibinfo {author}
  {\bibfnamefont {S.}~\bibnamefont {McBride}}, \bibinfo {author} {\bibfnamefont
  {A.}~\bibnamefont {Braun}}, \bibinfo {author} {\bibfnamefont
  {W.}~\bibnamefont {Lee}}, \bibinfo {author} {\bibfnamefont {V.}~\bibnamefont
  {Lucivero}},\ and\ \bibinfo {author} {\bibfnamefont {M.}~\bibnamefont
  {Romalis}},\ }\bibfield  {title} {\bibinfo {title} {Portable magnetometry for
  detection of biomagnetism in ambient environments},\ }\href
  {https://doi.org/10.1103/PhysRevApplied.14.011002} {\bibfield  {journal}
  {\bibinfo  {journal} {Phys. Rev. Applied}\ }\textbf {\bibinfo {volume}
  {14}},\ \bibinfo {pages} {011002} (\bibinfo {year} {2020})}\BibitemShut
  {NoStop}%
\bibitem [{\citenamefont {Lucivero}\ \emph {et~al.}(2021)\citenamefont
  {Lucivero}, \citenamefont {Lee}, \citenamefont {Dural},\ and\ \citenamefont
  {Romalis}}]{LuciveroPRAppl2021}%
  \BibitemOpen
  \bibfield  {author} {\bibinfo {author} {\bibfnamefont {V.}~\bibnamefont
  {Lucivero}}, \bibinfo {author} {\bibfnamefont {W.}~\bibnamefont {Lee}},
  \bibinfo {author} {\bibfnamefont {N.}~\bibnamefont {Dural}},\ and\ \bibinfo
  {author} {\bibfnamefont {M.}~\bibnamefont {Romalis}},\ }\bibfield  {title}
  {\bibinfo {title} {Femtotesla direct magnetic gradiometer using a single
  multipass cell},\ }\href {https://doi.org/10.1103/PhysRevApplied.15.014004}
  {\bibfield  {journal} {\bibinfo  {journal} {Phys. Rev. Applied}\ }\textbf
  {\bibinfo {volume} {15}},\ \bibinfo {pages} {014004} (\bibinfo {year}
  {2021})}\BibitemShut {NoStop}%
\bibitem [{\citenamefont {Gomez}\ \emph {et~al.}(2020)\citenamefont {Gomez},
  \citenamefont {Martin}, \citenamefont {Mazzinghi}, \citenamefont
  {Benedicto~Orenes}, \citenamefont {Palacios},\ and\ \citenamefont
  {Mitchell}}]{PhysRevLett.124.170401}%
  \BibitemOpen
  \bibfield  {author} {\bibinfo {author} {\bibfnamefont {P.}~\bibnamefont
  {Gomez}}, \bibinfo {author} {\bibfnamefont {F.}~\bibnamefont {Martin}},
  \bibinfo {author} {\bibfnamefont {C.}~\bibnamefont {Mazzinghi}}, \bibinfo
  {author} {\bibfnamefont {D.}~\bibnamefont {Benedicto~Orenes}}, \bibinfo
  {author} {\bibfnamefont {S.}~\bibnamefont {Palacios}},\ and\ \bibinfo
  {author} {\bibfnamefont {M.~W.}\ \bibnamefont {Mitchell}},\ }\bibfield
  {title} {\bibinfo {title} {Bose-einstein condensate comagnetometer},\ }\href
  {https://doi.org/10.1103/PhysRevLett.124.170401} {\bibfield  {journal}
  {\bibinfo  {journal} {Phys. Rev. Lett.}\ }\textbf {\bibinfo {volume} {124}},\
  \bibinfo {pages} {170401} (\bibinfo {year} {2020})}\BibitemShut {NoStop}%
\bibitem [{\citenamefont {Vasilakis}\ \emph {et~al.}(2009)\citenamefont
  {Vasilakis}, \citenamefont {Brown}, \citenamefont {Kornack},\ and\
  \citenamefont {Romalis}}]{RomalisVas-comag}%
  \BibitemOpen
  \bibfield  {author} {\bibinfo {author} {\bibfnamefont {G.}~\bibnamefont
  {Vasilakis}}, \bibinfo {author} {\bibfnamefont {J.~M.}\ \bibnamefont
  {Brown}}, \bibinfo {author} {\bibfnamefont {T.~W.}\ \bibnamefont {Kornack}},\
  and\ \bibinfo {author} {\bibfnamefont {M.~V.}\ \bibnamefont {Romalis}},\
  }\bibfield  {title} {\bibinfo {title} {Limits on new long range nuclear
  spin-dependent forces set with a
  $\mathbf{K}\mathrm{\text{\ensuremath{-}}}^{3}\mathrm{He}$ comagnetometer},\
  }\href {https://link.aps.org/doi/10.1103/PhysRevLett.103.261801} {\bibfield
  {journal} {\bibinfo  {journal} {Phys. Rev. Lett.}\ }\textbf {\bibinfo
  {volume} {\textbf{103}}},\ \bibinfo {pages} {261801} (\bibinfo {year}
  {2009})}\BibitemShut {NoStop}%
\bibitem [{\citenamefont {Walker}\ and\ \citenamefont
  {Larsen}(2016)}]{WalkerBook2016}%
  \BibitemOpen
  \bibfield  {author} {\bibinfo {author} {\bibfnamefont {T.}~\bibnamefont
  {Walker}}\ and\ \bibinfo {author} {\bibfnamefont {M.}~\bibnamefont
  {Larsen}},\ }\bibfield  {title} {\bibinfo {title} {Spin-exchange-pumped {NMR}
  gyros},\ }in\ \href {https://doi.org/10.1016/bs.aamop.2016.04.002} {\emph
  {\bibinfo {booktitle} {Advances In Atomic, Molecular, and Optical Physics}}}\
  (\bibinfo  {publisher} {Elsevier},\ \bibinfo {year} {2016})\ pp.\ \bibinfo
  {pages} {373--401}\BibitemShut {NoStop}%
\bibitem [{\citenamefont {Fang}\ and\ \citenamefont
  {Qin}(2012)}]{FangSensors2012}%
  \BibitemOpen
  \bibfield  {author} {\bibinfo {author} {\bibfnamefont {J.}~\bibnamefont
  {Fang}}\ and\ \bibinfo {author} {\bibfnamefont {J.}~\bibnamefont {Qin}},\
  }\bibfield  {title} {\bibinfo {title} {Advances in atomic gyroscopes: A view
  from inertial navigation applications},\ }\href
  {https://doi.org/10.3390/s120506331} {\bibfield  {journal} {\bibinfo
  {journal} {Sensors}\ }\textbf {\bibinfo {volume} {12}},\ \bibinfo {pages}
  {6331} (\bibinfo {year} {2012})}\BibitemShut {NoStop}%
\bibitem [{\citenamefont {Shaham}\ \emph {et~al.}(2022)\citenamefont {Shaham},
  \citenamefont {Katz},\ and\ \citenamefont {Firstenberg}}]{Shaham2022}%
  \BibitemOpen
  \bibfield  {author} {\bibinfo {author} {\bibfnamefont {R.}~\bibnamefont
  {Shaham}}, \bibinfo {author} {\bibfnamefont {O.}~\bibnamefont {Katz}},\ and\
  \bibinfo {author} {\bibfnamefont {O.}~\bibnamefont {Firstenberg}},\
  }\bibfield  {title} {\bibinfo {title} {Strong coupling of alkali-metal spins
  to noble-gas spins with an hour-long coherence time},\ }\bibfield  {journal}
  {\bibinfo  {journal} {Nature Physics}\ }\href
  {https://doi.org/10.1038/s41567-022-01535-w} {10.1038/s41567-022-01535-w}
  (\bibinfo {year} {2022})\BibitemShut {NoStop}%
\bibitem [{\citenamefont {Afach}\ \emph {et~al.}(2021)\citenamefont {Afach},
  \citenamefont {Buchler}, \citenamefont {Budker}, \citenamefont {Dailey},
  \citenamefont {Derevianko}, \citenamefont {Dumont}, \citenamefont {Figueroa},
  \citenamefont {Gerhardt}, \citenamefont {Gruji{\'{c}}}, \citenamefont {Guo},
  \citenamefont {Hao}, \citenamefont {Hamilton}, \citenamefont {Hedges},
  \citenamefont {Kimball}, \citenamefont {Kim}, \citenamefont {Khamis},
  \citenamefont {Kornack}, \citenamefont {Lebedev}, \citenamefont {Lu},
  \citenamefont {Masia-Roig}, \citenamefont {Monroy}, \citenamefont {Padniuk},
  \citenamefont {Palm}, \citenamefont {Park}, \citenamefont {Paul},
  \citenamefont {Penaflor}, \citenamefont {Peng}, \citenamefont {Pospelov},
  \citenamefont {Preston}, \citenamefont {Pustelny}, \citenamefont {Scholtes},
  \citenamefont {Segura}, \citenamefont {Semertzidis}, \citenamefont {Sheng},
  \citenamefont {Shin}, \citenamefont {Smiga}, \citenamefont {Stalnaker},
  \citenamefont {Sulai}, \citenamefont {Tandon}, \citenamefont {Wang},
  \citenamefont {Weis}, \citenamefont {Wickenbrock}, \citenamefont {Wilson},
  \citenamefont {Wu}, \citenamefont {Wurm}, \citenamefont {Xiao}, \citenamefont
  {Yang}, \citenamefont {Yu},\ and\ \citenamefont {Zhang}}]{Afach2021}%
  \BibitemOpen
  \bibfield  {author} {\bibinfo {author} {\bibfnamefont {S.}~\bibnamefont
  {Afach}}, \bibinfo {author} {\bibfnamefont {B.~C.}\ \bibnamefont {Buchler}},
  \bibinfo {author} {\bibfnamefont {D.}~\bibnamefont {Budker}}, \bibinfo
  {author} {\bibfnamefont {C.}~\bibnamefont {Dailey}}, \bibinfo {author}
  {\bibfnamefont {A.}~\bibnamefont {Derevianko}}, \bibinfo {author}
  {\bibfnamefont {V.}~\bibnamefont {Dumont}}, \bibinfo {author} {\bibfnamefont
  {N.~L.}\ \bibnamefont {Figueroa}}, \bibinfo {author} {\bibfnamefont
  {I.}~\bibnamefont {Gerhardt}}, \bibinfo {author} {\bibfnamefont {Z.~D.}\
  \bibnamefont {Gruji{\'{c}}}}, \bibinfo {author} {\bibfnamefont
  {H.}~\bibnamefont {Guo}}, \bibinfo {author} {\bibfnamefont {C.}~\bibnamefont
  {Hao}}, \bibinfo {author} {\bibfnamefont {P.~S.}\ \bibnamefont {Hamilton}},
  \bibinfo {author} {\bibfnamefont {M.}~\bibnamefont {Hedges}}, \bibinfo
  {author} {\bibfnamefont {D.~F.~J.}\ \bibnamefont {Kimball}}, \bibinfo
  {author} {\bibfnamefont {D.}~\bibnamefont {Kim}}, \bibinfo {author}
  {\bibfnamefont {S.}~\bibnamefont {Khamis}}, \bibinfo {author} {\bibfnamefont
  {T.}~\bibnamefont {Kornack}}, \bibinfo {author} {\bibfnamefont
  {V.}~\bibnamefont {Lebedev}}, \bibinfo {author} {\bibfnamefont {Z.-T.}\
  \bibnamefont {Lu}}, \bibinfo {author} {\bibfnamefont {H.}~\bibnamefont
  {Masia-Roig}}, \bibinfo {author} {\bibfnamefont {M.}~\bibnamefont {Monroy}},
  \bibinfo {author} {\bibfnamefont {M.}~\bibnamefont {Padniuk}}, \bibinfo
  {author} {\bibfnamefont {C.~A.}\ \bibnamefont {Palm}}, \bibinfo {author}
  {\bibfnamefont {S.~Y.}\ \bibnamefont {Park}}, \bibinfo {author}
  {\bibfnamefont {K.~V.}\ \bibnamefont {Paul}}, \bibinfo {author}
  {\bibfnamefont {A.}~\bibnamefont {Penaflor}}, \bibinfo {author}
  {\bibfnamefont {X.}~\bibnamefont {Peng}}, \bibinfo {author} {\bibfnamefont
  {M.}~\bibnamefont {Pospelov}}, \bibinfo {author} {\bibfnamefont
  {R.}~\bibnamefont {Preston}}, \bibinfo {author} {\bibfnamefont
  {S.}~\bibnamefont {Pustelny}}, \bibinfo {author} {\bibfnamefont
  {T.}~\bibnamefont {Scholtes}}, \bibinfo {author} {\bibfnamefont {P.~C.}\
  \bibnamefont {Segura}}, \bibinfo {author} {\bibfnamefont {Y.~K.}\
  \bibnamefont {Semertzidis}}, \bibinfo {author} {\bibfnamefont
  {D.}~\bibnamefont {Sheng}}, \bibinfo {author} {\bibfnamefont {Y.~C.}\
  \bibnamefont {Shin}}, \bibinfo {author} {\bibfnamefont {J.~A.}\ \bibnamefont
  {Smiga}}, \bibinfo {author} {\bibfnamefont {J.~E.}\ \bibnamefont
  {Stalnaker}}, \bibinfo {author} {\bibfnamefont {I.}~\bibnamefont {Sulai}},
  \bibinfo {author} {\bibfnamefont {D.}~\bibnamefont {Tandon}}, \bibinfo
  {author} {\bibfnamefont {T.}~\bibnamefont {Wang}}, \bibinfo {author}
  {\bibfnamefont {A.}~\bibnamefont {Weis}}, \bibinfo {author} {\bibfnamefont
  {A.}~\bibnamefont {Wickenbrock}}, \bibinfo {author} {\bibfnamefont
  {T.}~\bibnamefont {Wilson}}, \bibinfo {author} {\bibfnamefont
  {T.}~\bibnamefont {Wu}}, \bibinfo {author} {\bibfnamefont {D.}~\bibnamefont
  {Wurm}}, \bibinfo {author} {\bibfnamefont {W.}~\bibnamefont {Xiao}}, \bibinfo
  {author} {\bibfnamefont {Y.}~\bibnamefont {Yang}}, \bibinfo {author}
  {\bibfnamefont {D.}~\bibnamefont {Yu}},\ and\ \bibinfo {author}
  {\bibfnamefont {J.}~\bibnamefont {Zhang}},\ }\bibfield  {title} {\bibinfo
  {title} {Search for topological defect dark matter with a global network of
  optical magnetometers},\ }\href {https://doi.org/10.1038/s41567-021-01393-y}
  {\bibfield  {journal} {\bibinfo  {journal} {Nature Physics}\ }\textbf
  {\bibinfo {volume} {17}},\ \bibinfo {pages} {1396} (\bibinfo {year}
  {2021})}\BibitemShut {NoStop}%
\bibitem [{\citenamefont {Almasi}\ \emph {et~al.}(2020)\citenamefont {Almasi},
  \citenamefont {Lee}, \citenamefont {Winarto}, \citenamefont {Smiciklas},\
  and\ \citenamefont {Romalis}}]{PhysRevLett.125.201802}%
  \BibitemOpen
  \bibfield  {author} {\bibinfo {author} {\bibfnamefont {A.}~\bibnamefont
  {Almasi}}, \bibinfo {author} {\bibfnamefont {J.}~\bibnamefont {Lee}},
  \bibinfo {author} {\bibfnamefont {H.}~\bibnamefont {Winarto}}, \bibinfo
  {author} {\bibfnamefont {M.}~\bibnamefont {Smiciklas}},\ and\ \bibinfo
  {author} {\bibfnamefont {M.~V.}\ \bibnamefont {Romalis}},\ }\bibfield
  {title} {\bibinfo {title} {New limits on anomalous spin-spin interactions},\
  }\href {https://doi.org/10.1103/PhysRevLett.125.201802} {\bibfield  {journal}
  {\bibinfo  {journal} {Phys. Rev. Lett.}\ }\textbf {\bibinfo {volume} {125}},\
  \bibinfo {pages} {201802} (\bibinfo {year} {2020})}\BibitemShut {NoStop}%
\bibitem [{\citenamefont {Pustelny}\ \emph {et~al.}(2013)\citenamefont
  {Pustelny}, \citenamefont {Jackson~Kimball}, \citenamefont {Pankow},
  \citenamefont {Ledbetter}, \citenamefont {Wlodarczyk}, \citenamefont
  {Wcislo}, \citenamefont {Pospelov}, \citenamefont {Smith}, \citenamefont
  {Read}, \citenamefont {Gawlik} \emph {et~al.}}]{pustelny2013global}%
  \BibitemOpen
  \bibfield  {author} {\bibinfo {author} {\bibfnamefont {S.}~\bibnamefont
  {Pustelny}}, \bibinfo {author} {\bibfnamefont {D.~F.}\ \bibnamefont
  {Jackson~Kimball}}, \bibinfo {author} {\bibfnamefont {C.}~\bibnamefont
  {Pankow}}, \bibinfo {author} {\bibfnamefont {M.~P.}\ \bibnamefont
  {Ledbetter}}, \bibinfo {author} {\bibfnamefont {P.}~\bibnamefont
  {Wlodarczyk}}, \bibinfo {author} {\bibfnamefont {P.}~\bibnamefont {Wcislo}},
  \bibinfo {author} {\bibfnamefont {M.}~\bibnamefont {Pospelov}}, \bibinfo
  {author} {\bibfnamefont {J.~R.}\ \bibnamefont {Smith}}, \bibinfo {author}
  {\bibfnamefont {J.}~\bibnamefont {Read}}, \bibinfo {author} {\bibfnamefont
  {W.}~\bibnamefont {Gawlik}}, \emph {et~al.},\ }\bibfield  {title} {\bibinfo
  {title} {The global network of optical magnetometers for exotic physics
  (gnome): A novel scheme to search for physics beyond the standard model},\
  }\href@noop {} {\bibfield  {journal} {\bibinfo  {journal} {Annalen der
  Physik}\ }\textbf {\bibinfo {volume} {525}},\ \bibinfo {pages} {659}
  (\bibinfo {year} {2013})}\BibitemShut {NoStop}%
\bibitem [{\citenamefont {Kong}\ \emph {et~al.}(2020)\citenamefont {Kong},
  \citenamefont {Jim{\'{e}}nez-Mart{\'{\i}}nez}, \citenamefont {Troullinou},
  \citenamefont {Lucivero}, \citenamefont {T{\'{o}}th},\ and\ \citenamefont
  {Mitchell}}]{MitchellNatureCom}%
  \BibitemOpen
  \bibfield  {author} {\bibinfo {author} {\bibfnamefont {J.}~\bibnamefont
  {Kong}}, \bibinfo {author} {\bibfnamefont {R.}~\bibnamefont
  {Jim{\'{e}}nez-Mart{\'{\i}}nez}}, \bibinfo {author} {\bibfnamefont
  {C.}~\bibnamefont {Troullinou}}, \bibinfo {author} {\bibfnamefont {V.~G.}\
  \bibnamefont {Lucivero}}, \bibinfo {author} {\bibfnamefont {G.}~\bibnamefont
  {T{\'{o}}th}},\ and\ \bibinfo {author} {\bibfnamefont {M.~W.}\ \bibnamefont
  {Mitchell}},\ }\bibfield  {title} {\bibinfo {title} {Measurement-induced,
  spatially-extended entanglement in a hot, strongly-interacting atomic
  system},\ }\bibfield  {journal} {\bibinfo  {journal} {Nature Communications}\
  }\textbf {\bibinfo {volume} {\textbf{11}}},\ \href
  {https://doi.org/10.1038/s41467-020-15899-1} {10.1038/s41467-020-15899-1}
  (\bibinfo {year} {2020})\BibitemShut {NoStop}%
\bibitem [{\citenamefont {Kuzmich}\ \emph {et~al.}(2000)\citenamefont
  {Kuzmich}, \citenamefont {Mandel},\ and\ \citenamefont
  {Bigelow}}]{KuzmichSpinSqueezingQND}%
  \BibitemOpen
  \bibfield  {author} {\bibinfo {author} {\bibfnamefont {A.}~\bibnamefont
  {Kuzmich}}, \bibinfo {author} {\bibfnamefont {L.}~\bibnamefont {Mandel}},\
  and\ \bibinfo {author} {\bibfnamefont {N.~P.}\ \bibnamefont {Bigelow}},\
  }\bibfield  {title} {\bibinfo {title} {Generation of spin squeezing via
  continuous quantum nondemolition measurement},\ }\href
  {https://doi.org/10.1103/PhysRevLett.85.1594} {\bibfield  {journal} {\bibinfo
   {journal} {Phys. Rev. Lett.}\ }\textbf {\bibinfo {volume} {85}},\ \bibinfo
  {pages} {1594} (\bibinfo {year} {2000})}\BibitemShut {NoStop}%
\bibitem [{\citenamefont {Vasilakis}\ \emph {et~al.}(2015)\citenamefont
  {Vasilakis}, \citenamefont {Shen}, \citenamefont {Jensen}, \citenamefont
  {Balabas}, \citenamefont {Salart}, \citenamefont {Chen},\ and\ \citenamefont
  {Polzik}}]{VasilakisPolzik2015}%
  \BibitemOpen
  \bibfield  {author} {\bibinfo {author} {\bibfnamefont {G.}~\bibnamefont
  {Vasilakis}}, \bibinfo {author} {\bibfnamefont {H.}~\bibnamefont {Shen}},
  \bibinfo {author} {\bibfnamefont {K.}~\bibnamefont {Jensen}}, \bibinfo
  {author} {\bibfnamefont {M.}~\bibnamefont {Balabas}}, \bibinfo {author}
  {\bibfnamefont {D.}~\bibnamefont {Salart}}, \bibinfo {author} {\bibfnamefont
  {B.}~\bibnamefont {Chen}},\ and\ \bibinfo {author} {\bibfnamefont {E.~S.}\
  \bibnamefont {Polzik}},\ }\bibfield  {title} {\bibinfo {title} {Generation of
  a squeezed state of an oscillator by stroboscopic back-action-evading
  measurement},\ }\href {https://doi.org/10.1038/nphys3280} {\bibfield
  {journal} {\bibinfo  {journal} {Nature Physics}\ }\textbf {\bibinfo {volume}
  {11}},\ \bibinfo {pages} {389} (\bibinfo {year} {2015})}\BibitemShut
  {NoStop}%
\bibitem [{\citenamefont {Shaham}\ \emph {et~al.}(2020)\citenamefont {Shaham},
  \citenamefont {Katz},\ and\ \citenamefont
  {Firstenberg}}]{PhysRevA.102.012822}%
  \BibitemOpen
  \bibfield  {author} {\bibinfo {author} {\bibfnamefont {R.}~\bibnamefont
  {Shaham}}, \bibinfo {author} {\bibfnamefont {O.}~\bibnamefont {Katz}},\ and\
  \bibinfo {author} {\bibfnamefont {O.}~\bibnamefont {Firstenberg}},\
  }\bibfield  {title} {\bibinfo {title} {Quantum dynamics of collective spin
  states in a thermal gas},\ }\href
  {https://doi.org/10.1103/PhysRevA.102.012822} {\bibfield  {journal} {\bibinfo
   {journal} {Phys. Rev. A}\ }\textbf {\bibinfo {volume} {102}},\ \bibinfo
  {pages} {012822} (\bibinfo {year} {2020})}\BibitemShut {NoStop}%
\bibitem [{\citenamefont {Katz}\ \emph {et~al.}(2022)\citenamefont {Katz},
  \citenamefont {Shaham},\ and\ \citenamefont
  {Firstenberg}}]{PRXQuantum.3.010305}%
  \BibitemOpen
  \bibfield  {author} {\bibinfo {author} {\bibfnamefont {O.}~\bibnamefont
  {Katz}}, \bibinfo {author} {\bibfnamefont {R.}~\bibnamefont {Shaham}},\ and\
  \bibinfo {author} {\bibfnamefont {O.}~\bibnamefont {Firstenberg}},\
  }\bibfield  {title} {\bibinfo {title} {Quantum interface for noble-gas spins
  based on spin-exchange collisions},\ }\href
  {https://doi.org/10.1103/PRXQuantum.3.010305} {\bibfield  {journal} {\bibinfo
   {journal} {PRX Quantum}\ }\textbf {\bibinfo {volume} {3}},\ \bibinfo {pages}
  {010305} (\bibinfo {year} {2022})}\BibitemShut {NoStop}%
\bibitem [{\citenamefont {Troullinou}\ \emph {et~al.}(2021)\citenamefont
  {Troullinou}, \citenamefont {Jim\'enez-Mart\'{\i}nez}, \citenamefont {Kong},
  \citenamefont {Lucivero},\ and\ \citenamefont {Mitchell}}]{Troullinou}%
  \BibitemOpen
  \bibfield  {author} {\bibinfo {author} {\bibfnamefont {C.}~\bibnamefont
  {Troullinou}}, \bibinfo {author} {\bibfnamefont {R.}~\bibnamefont
  {Jim\'enez-Mart\'{\i}nez}}, \bibinfo {author} {\bibfnamefont
  {J.}~\bibnamefont {Kong}}, \bibinfo {author} {\bibfnamefont {V.~G.}\
  \bibnamefont {Lucivero}},\ and\ \bibinfo {author} {\bibfnamefont {M.~W.}\
  \bibnamefont {Mitchell}},\ }\bibfield  {title} {\bibinfo {title}
  {Squeezed-light enhancement and backaction evasion in a high sensitivity
  optically pumped magnetometer},\ }\href
  {https://doi.org/10.1103/PhysRevLett.127.193601} {\bibfield  {journal}
  {\bibinfo  {journal} {Phys. Rev. Lett.}\ }\textbf {\bibinfo {volume} {127}},\
  \bibinfo {pages} {193601} (\bibinfo {year} {2021})}\BibitemShut {NoStop}%
\bibitem [{\citenamefont {Julsgaard}\ \emph {et~al.}(2001)\citenamefont
  {Julsgaard}, \citenamefont {Kozhekin},\ and\ \citenamefont
  {Polzik}}]{Julsgaard2001}%
  \BibitemOpen
  \bibfield  {author} {\bibinfo {author} {\bibfnamefont {B.}~\bibnamefont
  {Julsgaard}}, \bibinfo {author} {\bibfnamefont {A.}~\bibnamefont
  {Kozhekin}},\ and\ \bibinfo {author} {\bibfnamefont {E.~S.}\ \bibnamefont
  {Polzik}},\ }\bibfield  {title} {\bibinfo {title} {Experimental long-lived
  entanglement of two macroscopic objects},\ }\href
  {https://doi.org/10.1038/35096524} {\bibfield  {journal} {\bibinfo  {journal}
  {Nature}\ }\textbf {\bibinfo {volume} {413}},\ \bibinfo {pages} {400}
  (\bibinfo {year} {2001})}\BibitemShut {NoStop}%
\bibitem [{\citenamefont {Sherson}\ \emph {et~al.}(2006)\citenamefont
  {Sherson}, \citenamefont {Krauter}, \citenamefont {Olsson}, \citenamefont
  {Julsgaard}, \citenamefont {Hammerer}, \citenamefont {Cirac},\ and\
  \citenamefont {Polzik}}]{Sherson2006}%
  \BibitemOpen
  \bibfield  {author} {\bibinfo {author} {\bibfnamefont {J.~F.}\ \bibnamefont
  {Sherson}}, \bibinfo {author} {\bibfnamefont {H.}~\bibnamefont {Krauter}},
  \bibinfo {author} {\bibfnamefont {R.~K.}\ \bibnamefont {Olsson}}, \bibinfo
  {author} {\bibfnamefont {B.}~\bibnamefont {Julsgaard}}, \bibinfo {author}
  {\bibfnamefont {K.}~\bibnamefont {Hammerer}}, \bibinfo {author}
  {\bibfnamefont {I.}~\bibnamefont {Cirac}},\ and\ \bibinfo {author}
  {\bibfnamefont {E.~S.}\ \bibnamefont {Polzik}},\ }\bibfield  {title}
  {\bibinfo {title} {Quantum teleportation between light and matter},\ }\href
  {https://doi.org/10.1038/nature05136} {\bibfield  {journal} {\bibinfo
  {journal} {Nature}\ }\textbf {\bibinfo {volume} {443}},\ \bibinfo {pages}
  {557} (\bibinfo {year} {2006})}\BibitemShut {NoStop}%
\bibitem [{\citenamefont {Serafin}\ \emph {et~al.}(2021)\citenamefont
  {Serafin}, \citenamefont {Fadel}, \citenamefont {Treutlein},\ and\
  \citenamefont {Sinatra}}]{TrutleinHe3}%
  \BibitemOpen
  \bibfield  {author} {\bibinfo {author} {\bibfnamefont {A.}~\bibnamefont
  {Serafin}}, \bibinfo {author} {\bibfnamefont {M.}~\bibnamefont {Fadel}},
  \bibinfo {author} {\bibfnamefont {P.}~\bibnamefont {Treutlein}},\ and\
  \bibinfo {author} {\bibfnamefont {A.}~\bibnamefont {Sinatra}},\ }\bibfield
  {title} {\bibinfo {title} {Nuclear spin squeezing in $\mathrm{Helium}$-3 by
  continuous quantum nondemolition measurement},\ }\href
  {https://doi.org/10.1103/PhysRevLett.127.013601} {\bibfield  {journal}
  {\bibinfo  {journal} {Phys. Rev. Lett.}\ }\textbf {\bibinfo {volume}
  {\textbf{127}}},\ \bibinfo {pages} {013601} (\bibinfo {year}
  {2021})}\BibitemShut {NoStop}%
\bibitem [{\citenamefont {Seltzer}(2008)}]{seltzerthesis}%
  \BibitemOpen
  \bibfield  {author} {\bibinfo {author} {\bibfnamefont {S.~J.}\ \bibnamefont
  {Seltzer}},\ }\bibfield  {title} {\bibinfo {title} {Developments in
  alkali-metal atomic magnetometry},\ }\href@noop {} {\  (\bibinfo {year}
  {2008})}\BibitemShut {NoStop}%
\bibitem [{\citenamefont {Savukov}\ and\ \citenamefont
  {Romalis}(2005)}]{SavukovRomalis}%
  \BibitemOpen
  \bibfield  {author} {\bibinfo {author} {\bibfnamefont {I.~M.}\ \bibnamefont
  {Savukov}}\ and\ \bibinfo {author} {\bibfnamefont {M.~V.}\ \bibnamefont
  {Romalis}},\ }\bibfield  {title} {\bibinfo {title} {Effects of spin-exchange
  collisions in a high-density alkali-metal vapor in low magnetic fields},\
  }\href {https://link.aps.org/doi/10.1103/PhysRevA.71.023405} {\bibfield
  {journal} {\bibinfo  {journal} {Phys. Rev. A}\ }\textbf {\bibinfo {volume}
  {71}},\ \bibinfo {pages} {023405} (\bibinfo {year} {2005})}\BibitemShut
  {NoStop}%
\bibitem [{\citenamefont {Grosset{\^{e}}te}(1964)}]{Grossette1}%
  \BibitemOpen
  \bibfield  {author} {\bibinfo {author} {\bibfnamefont {F.}~\bibnamefont
  {Grosset{\^{e}}te}},\ }\bibfield  {title} {\bibinfo {title} {Relaxation par
  collisions d{\textquotesingle}{\'{e}}change de spin},\ }\href
  {https://doi.org/10.1051/jphys:01964002504038300} {\bibfield  {journal}
  {\bibinfo  {journal} {Journal de Physique}\ }\textbf {\bibinfo {volume}
  {25}},\ \bibinfo {pages} {383} (\bibinfo {year} {1964})}\BibitemShut
  {NoStop}%
\bibitem [{\citenamefont {Grosset{\^{e}}te}(1968)}]{Grossette2}%
  \BibitemOpen
  \bibfield  {author} {\bibinfo {author} {\bibfnamefont {F.}~\bibnamefont
  {Grosset{\^{e}}te}},\ }\bibfield  {title} {\bibinfo {title} {Relaxation par
  collisions d{\textquotesingle}{\'{e}}change de spins. ({II})},\ }\href
  {https://doi.org/10.1051/jphys:01968002905-6045600} {\bibfield  {journal}
  {\bibinfo  {journal} {Journal de Physique}\ }\textbf {\bibinfo {volume}
  {29}},\ \bibinfo {pages} {456} (\bibinfo {year} {1968})}\BibitemShut
  {NoStop}%
\bibitem [{\citenamefont {Gibbs}\ and\ \citenamefont
  {Hull}(1967)}]{PhysRev.153.132}%
  \BibitemOpen
  \bibfield  {author} {\bibinfo {author} {\bibfnamefont {H.~M.}\ \bibnamefont
  {Gibbs}}\ and\ \bibinfo {author} {\bibfnamefont {R.~J.}\ \bibnamefont
  {Hull}},\ }\bibfield  {title} {\bibinfo {title} {Spin-exchange cross sections
  for $\mathbf{Rb}^{87}$-$\mathbf{Rb}^{87}$ and
  $\mathbf{Rb}^{87}$-$\mathbf{Cs}^{133}$ collisions},\ }\href
  {https://link.aps.org/doi/10.1103/PhysRev.153.132} {\bibfield  {journal}
  {\bibinfo  {journal} {Phys. Rev.}\ }\textbf {\bibinfo {volume}
  {\textbf{153}}} (\bibinfo {year} {1967})}\BibitemShut {NoStop}%
\bibitem [{\citenamefont {Happer}\ and\ \citenamefont
  {Tang}(1973)}]{PhysRevLett.31.273}%
  \BibitemOpen
  \bibfield  {author} {\bibinfo {author} {\bibfnamefont {W.}~\bibnamefont
  {Happer}}\ and\ \bibinfo {author} {\bibfnamefont {H.}~\bibnamefont {Tang}},\
  }\bibfield  {title} {\bibinfo {title} {Spin-exchange shift and narrowing of
  magnetic resonance lines in optically pumped alkali vapors},\ }\href
  {https://link.aps.org/doi/10.1103/PhysRevLett.31.273} {\bibfield  {journal}
  {\bibinfo  {journal} {Phys. Rev. Lett.}\ }\textbf {\bibinfo {volume}
  {\textbf{31}}} (\bibinfo {year} {1973})}\BibitemShut {NoStop}%
\bibitem [{\citenamefont {Happer}\ and\ \citenamefont
  {Tam}(1977)}]{PhysRevA.16.1877}%
  \BibitemOpen
  \bibfield  {author} {\bibinfo {author} {\bibfnamefont {W.}~\bibnamefont
  {Happer}}\ and\ \bibinfo {author} {\bibfnamefont {A.~C.}\ \bibnamefont
  {Tam}},\ }\bibfield  {title} {\bibinfo {title} {Effect of rapid spin exchange
  on the magnetic-resonance spectrum of alkali vapors},\ }\href
  {https://doi.org/10.1103/PhysRevA.16.1877} {\bibfield  {journal} {\bibinfo
  {journal} {Phys. Rev. A}\ }\textbf {\bibinfo {volume} {\textbf{16}}},\
  \bibinfo {pages} {1877} (\bibinfo {year} {1977})}\BibitemShut {NoStop}%
\bibitem [{\citenamefont {Lucivero}\ \emph
  {et~al.}(2017{\natexlab{a}})\citenamefont {Lucivero}, \citenamefont {Dimic},
  \citenamefont {Kong}, \citenamefont {Jim\'enez-Mart\'{\i}nez},\ and\
  \citenamefont {Mitchell}}]{lucivero2017sensitivity}%
  \BibitemOpen
  \bibfield  {author} {\bibinfo {author} {\bibfnamefont {V.~G.}\ \bibnamefont
  {Lucivero}}, \bibinfo {author} {\bibfnamefont {A.}~\bibnamefont {Dimic}},
  \bibinfo {author} {\bibfnamefont {J.}~\bibnamefont {Kong}}, \bibinfo {author}
  {\bibfnamefont {R.}~\bibnamefont {Jim\'enez-Mart\'{\i}nez}},\ and\ \bibinfo
  {author} {\bibfnamefont {M.~W.}\ \bibnamefont {Mitchell}},\ }\bibfield
  {title} {\bibinfo {title} {Sensitivity, quantum limits, and quantum
  enhancement of noise spectroscopies},\ }\href
  {https://doi.org/10.1103/PhysRevA.95.041803} {\bibfield  {journal} {\bibinfo
  {journal} {Phys. Rev. A}\ }\textbf {\bibinfo {volume} {\textbf{95}}},\
  \bibinfo {pages} {041803} (\bibinfo {year} {2017}{\natexlab{a}})}\BibitemShut
  {NoStop}%
\bibitem [{\citenamefont {Wen}\ \emph {et~al.}(2021)\citenamefont {Wen},
  \citenamefont {Li}, \citenamefont {Zhang},\ and\ \citenamefont
  {Zhao}}]{PhysRevA.104.063708}%
  \BibitemOpen
  \bibfield  {author} {\bibinfo {author} {\bibfnamefont {Y.}~\bibnamefont
  {Wen}}, \bibinfo {author} {\bibfnamefont {X.}~\bibnamefont {Li}}, \bibinfo
  {author} {\bibfnamefont {G.}~\bibnamefont {Zhang}},\ and\ \bibinfo {author}
  {\bibfnamefont {K.}~\bibnamefont {Zhao}},\ }\bibfield  {title} {\bibinfo
  {title} {Zero-field spin-noise spectrum of an alkali vapor with strong
  spin-exchange coupling},\ }\href
  {https://doi.org/10.1103/PhysRevA.104.063708} {\bibfield  {journal} {\bibinfo
   {journal} {Phys. Rev. A}\ }\textbf {\bibinfo {volume} {104}},\ \bibinfo
  {pages} {063708} (\bibinfo {year} {2021})}\BibitemShut {NoStop}%
\bibitem [{\citenamefont {Bruun}\ \emph {et~al.}(2009)\citenamefont {Bruun},
  \citenamefont {Andersen}, \citenamefont {Demler},\ and\ \citenamefont
  {S\o{}rensen}}]{BruunPRL2009}%
  \BibitemOpen
  \bibfield  {author} {\bibinfo {author} {\bibfnamefont {G.~M.}\ \bibnamefont
  {Bruun}}, \bibinfo {author} {\bibfnamefont {B.~M.}\ \bibnamefont {Andersen}},
  \bibinfo {author} {\bibfnamefont {E.}~\bibnamefont {Demler}},\ and\ \bibinfo
  {author} {\bibfnamefont {A.~S.}\ \bibnamefont {S\o{}rensen}},\ }\bibfield
  {title} {\bibinfo {title} {Probing spatial spin correlations of ultracold
  gases by quantum noise spectroscopy},\ }\href
  {https://doi.org/10.1103/PhysRevLett.102.030401} {\bibfield  {journal}
  {\bibinfo  {journal} {Phys. Rev. Lett.}\ }\textbf {\bibinfo {volume} {102}},\
  \bibinfo {pages} {030401} (\bibinfo {year} {2009})}\BibitemShut {NoStop}%
\bibitem [{\citenamefont {M\"{u}ller}\ \emph {et~al.}(2010)\citenamefont
  {M\"{u}ller}, \citenamefont {R\"{o}mer}, \citenamefont {H\"{u}bner},\ and\
  \citenamefont {Oestreich}}]{MullerAPL2010}%
  \BibitemOpen
  \bibfield  {author} {\bibinfo {author} {\bibfnamefont {G.~M.}\ \bibnamefont
  {M\"{u}ller}}, \bibinfo {author} {\bibfnamefont {M.}~\bibnamefont
  {R\"{o}mer}}, \bibinfo {author} {\bibfnamefont {J.}~\bibnamefont
  {H\"{u}bner}},\ and\ \bibinfo {author} {\bibfnamefont {M.}~\bibnamefont
  {Oestreich}},\ }\bibfield  {title} {\bibinfo {title} {Efficient data
  averaging for spin noise spectroscopy in semiconductors},\ }\href
  {https://doi.org/10.1063/1.3505342} {\bibfield  {journal} {\bibinfo
  {journal} {Applied Physics Letters}\ }\textbf {\bibinfo {volume} {97}},\
  \bibinfo {pages} {192109} (\bibinfo {year} {2010})}\BibitemShut {NoStop}%
\bibitem [{\citenamefont {Katsoprinakis}\ \emph {et~al.}(2007)\citenamefont
  {Katsoprinakis}, \citenamefont {Dellis},\ and\ \citenamefont
  {Kominis}}]{katsoprinakis}%
  \BibitemOpen
  \bibfield  {author} {\bibinfo {author} {\bibfnamefont {G.~E.}\ \bibnamefont
  {Katsoprinakis}}, \bibinfo {author} {\bibfnamefont {A.~T.}\ \bibnamefont
  {Dellis}},\ and\ \bibinfo {author} {\bibfnamefont {I.~K.}\ \bibnamefont
  {Kominis}},\ }\bibfield  {title} {\bibinfo {title} {Measurement of transverse
  spin-relaxation rates in a rubidium vapor by use of spin-noise
  spectroscopy},\ }\href {https://doi.org/10.1103/PhysRevA.75.042502}
  {\bibfield  {journal} {\bibinfo  {journal} {Phys. Rev. A}\ }\textbf {\bibinfo
  {volume} {\textbf{75}}},\ \bibinfo {pages} {042502} (\bibinfo {year}
  {2007})}\BibitemShut {NoStop}%
\bibitem [{\citenamefont {Appelt}\ \emph {et~al.}(1998)\citenamefont {Appelt},
  \citenamefont {Baranga}, \citenamefont {Erickson}, \citenamefont {Romalis},
  \citenamefont {Young},\ and\ \citenamefont {Happer}}]{Appelt}%
  \BibitemOpen
  \bibfield  {author} {\bibinfo {author} {\bibfnamefont {S.}~\bibnamefont
  {Appelt}}, \bibinfo {author} {\bibfnamefont {A.~B.-A.}\ \bibnamefont
  {Baranga}}, \bibinfo {author} {\bibfnamefont {C.~J.}\ \bibnamefont
  {Erickson}}, \bibinfo {author} {\bibfnamefont {M.~V.}\ \bibnamefont
  {Romalis}}, \bibinfo {author} {\bibfnamefont {A.~R.}\ \bibnamefont {Young}},\
  and\ \bibinfo {author} {\bibfnamefont {W.}~\bibnamefont {Happer}},\
  }\bibfield  {title} {\bibinfo {title} {Theory of spin-exchange optical
  pumping of ${}^{3}\mathbf{He}$ and ${}^{129}\mathbf{Xe}$},\ }\href
  {https://doi.org/10.1103/PhysRevA.58.1412} {\bibfield  {journal} {\bibinfo
  {journal} {Phys. Rev. A}\ }\textbf {\bibinfo {volume} {\textbf{58}}},\
  \bibinfo {pages} {1412} (\bibinfo {year} {1998})}\BibitemShut {NoStop}%
\bibitem [{\citenamefont {Happer}(1972)}]{HapperRev}%
  \BibitemOpen
  \bibfield  {author} {\bibinfo {author} {\bibfnamefont {W.}~\bibnamefont
  {Happer}},\ }\bibfield  {title} {\bibinfo {title} {Optical pumping},\ }\href
  {https://doi.org/10.1103/RevModPhys.44.169} {\bibfield  {journal} {\bibinfo
  {journal} {Rev. Mod. Phys.}\ }\textbf {\bibinfo {volume} {\textbf{44}}},\
  \bibinfo {pages} {169} (\bibinfo {year} {1972})}\BibitemShut {NoStop}%
\bibitem [{\citenamefont {Budker}\ and\ \citenamefont
  {Kimball}(2013)}]{optical_mag}%
  \BibitemOpen
  \bibfield  {author} {\bibinfo {author} {\bibfnamefont {D.}~\bibnamefont
  {Budker}}\ and\ \bibinfo {author} {\bibfnamefont {D.~F.}\ \bibnamefont
  {Kimball}},\ }\href@noop {} {\emph {\bibinfo {title} {Optical
  Magnetometry}}}\ (\bibinfo  {publisher} {Cambridge University Press},\
  \bibinfo {year} {2013})\BibitemShut {NoStop}%
\bibitem [{\citenamefont {Xiao}\ \emph {et~al.}(2021)\citenamefont {Xiao},
  \citenamefont {Wu}, \citenamefont {Peng},\ and\ \citenamefont
  {Guo}}]{PhysRevA.103.043116}%
  \BibitemOpen
  \bibfield  {author} {\bibinfo {author} {\bibfnamefont {W.}~\bibnamefont
  {Xiao}}, \bibinfo {author} {\bibfnamefont {T.}~\bibnamefont {Wu}}, \bibinfo
  {author} {\bibfnamefont {X.}~\bibnamefont {Peng}},\ and\ \bibinfo {author}
  {\bibfnamefont {H.}~\bibnamefont {Guo}},\ }\bibfield  {title} {\bibinfo
  {title} {Atomic spin-exchange collisions in magnetic fields},\ }\href
  {https://doi.org/10.1103/PhysRevA.103.043116} {\bibfield  {journal} {\bibinfo
   {journal} {Phys. Rev. A}\ }\textbf {\bibinfo {volume} {103}},\ \bibinfo
  {pages} {043116} (\bibinfo {year} {2021})}\BibitemShut {NoStop}%
\bibitem [{\citenamefont {Katz}\ \emph {et~al.}(2015)\citenamefont {Katz},
  \citenamefont {Peleg},\ and\ \citenamefont {Firstenberg}}]{katz2015coherent}%
  \BibitemOpen
  \bibfield  {author} {\bibinfo {author} {\bibfnamefont {O.}~\bibnamefont
  {Katz}}, \bibinfo {author} {\bibfnamefont {O.}~\bibnamefont {Peleg}},\ and\
  \bibinfo {author} {\bibfnamefont {O.}~\bibnamefont {Firstenberg}},\
  }\bibfield  {title} {\bibinfo {title} {Coherent coupling of alkali atoms by
  random collisions},\ }\href {https://doi.org/10.1103/PhysRevLett.115.113003}
  {\bibfield  {journal} {\bibinfo  {journal} {Phys. Rev. Lett.}\ }\textbf
  {\bibinfo {volume} {\textbf{115}}},\ \bibinfo {pages} {113003} (\bibinfo
  {year} {2015})}\BibitemShut {NoStop}%
\bibitem [{\citenamefont {Katz}\ \emph {et~al.}(2013)\citenamefont {Katz},
  \citenamefont {Dikopoltsev}, \citenamefont {Peleg}, \citenamefont {Shuker},
  \citenamefont {Steinhauer},\ and\ \citenamefont
  {Katz}}]{PhysRevLett.110.263004}%
  \BibitemOpen
  \bibfield  {author} {\bibinfo {author} {\bibfnamefont {O.}~\bibnamefont
  {Katz}}, \bibinfo {author} {\bibfnamefont {M.}~\bibnamefont {Dikopoltsev}},
  \bibinfo {author} {\bibfnamefont {O.}~\bibnamefont {Peleg}}, \bibinfo
  {author} {\bibfnamefont {M.}~\bibnamefont {Shuker}}, \bibinfo {author}
  {\bibfnamefont {J.}~\bibnamefont {Steinhauer}},\ and\ \bibinfo {author}
  {\bibfnamefont {N.}~\bibnamefont {Katz}},\ }\bibfield  {title} {\bibinfo
  {title} {Nonlinear elimination of spin-exchange relaxation of high magnetic
  moments},\ }\href {https://doi.org/10.1103/PhysRevLett.110.263004} {\bibfield
   {journal} {\bibinfo  {journal} {Phys. Rev. Lett.}\ }\textbf {\bibinfo
  {volume} {110}},\ \bibinfo {pages} {263004} (\bibinfo {year}
  {2013})}\BibitemShut {NoStop}%
\bibitem [{\citenamefont {Katz}\ and\ \citenamefont
  {Firstenberg}(2018)}]{ofer}%
  \BibitemOpen
  \bibfield  {author} {\bibinfo {author} {\bibfnamefont {O.}~\bibnamefont
  {Katz}}\ and\ \bibinfo {author} {\bibfnamefont {O.}~\bibnamefont
  {Firstenberg}},\ }\bibfield  {title} {\bibinfo {title} {Synchronization of
  strongly interacting alkali-metal spins},\ }\href
  {https://link.aps.org/doi/10.1103/PhysRevA.98.012712} {\bibfield  {journal}
  {\bibinfo  {journal} {Phys. Rev. A}\ }\textbf {\bibinfo {volume}
  {\textbf{98}}},\ \bibinfo {pages} {012712} (\bibinfo {year}
  {2018})}\BibitemShut {NoStop}%
\bibitem [{\citenamefont {Jim\'enez-Mart\'{\i}nez}\ \emph
  {et~al.}(2018)\citenamefont {Jim\'enez-Mart\'{\i}nez}, \citenamefont
  {Ko\l{}ody\ifmmode~\acute{n}\else \'{n}\fi{}ski}, \citenamefont {Troullinou},
  \citenamefont {Lucivero}, \citenamefont {Kong},\ and\ \citenamefont
  {Mitchell}}]{PhysRevLett.120.040503}%
  \BibitemOpen
  \bibfield  {author} {\bibinfo {author} {\bibfnamefont {R.}~\bibnamefont
  {Jim\'enez-Mart\'{\i}nez}}, \bibinfo {author} {\bibfnamefont
  {J.}~\bibnamefont {Ko\l{}ody\ifmmode~\acute{n}\else \'{n}\fi{}ski}}, \bibinfo
  {author} {\bibfnamefont {C.}~\bibnamefont {Troullinou}}, \bibinfo {author}
  {\bibfnamefont {V.~G.}\ \bibnamefont {Lucivero}}, \bibinfo {author}
  {\bibfnamefont {J.}~\bibnamefont {Kong}},\ and\ \bibinfo {author}
  {\bibfnamefont {M.~W.}\ \bibnamefont {Mitchell}},\ }\bibfield  {title}
  {\bibinfo {title} {Signal tracking beyond the time resolution of an atomic
  sensor by kalman filtering},\ }\href
  {https://doi.org/10.1103/PhysRevLett.120.040503} {\bibfield  {journal}
  {\bibinfo  {journal} {Phys. Rev. Lett.}\ }\textbf {\bibinfo {volume} {120}},\
  \bibinfo {pages} {040503} (\bibinfo {year} {2018})}\BibitemShut {NoStop}%
\bibitem [{Note1()}]{Note1}%
  \BibitemOpen
  \bibinfo {note} {An alternative approach, considering the dynamical matrix
  $A$ to be stochastic, and thus describing the spin dynamics with a random
  ordinary differential equation (RODE), appears to be equivalent to this
  method \cite {HanBook2017}}\BibitemShut {NoStop}%
\bibitem [{\citenamefont {Gardiner}(2009)}]{gardiner2009stochastic}%
  \BibitemOpen
  \bibfield  {author} {\bibinfo {author} {\bibfnamefont {C.}~\bibnamefont
  {Gardiner}},\ }\href@noop {} {\emph {\bibinfo {title} {Stochastic
  methods}}},\ Vol.\ \bibinfo {volume} {\textbf{4}}\ (\bibinfo  {publisher}
  {Springer Berlin},\ \bibinfo {year} {2009})\BibitemShut {NoStop}%
\bibitem [{\citenamefont {Shah}\ \emph {et~al.}(2010)\citenamefont {Shah},
  \citenamefont {Vasilakis},\ and\ \citenamefont
  {Romalis}}]{vasilakis2011stroboscopic}%
  \BibitemOpen
  \bibfield  {author} {\bibinfo {author} {\bibfnamefont {V.}~\bibnamefont
  {Shah}}, \bibinfo {author} {\bibfnamefont {G.}~\bibnamefont {Vasilakis}},\
  and\ \bibinfo {author} {\bibfnamefont {M.~V.}\ \bibnamefont {Romalis}},\
  }\bibfield  {title} {\bibinfo {title} {High bandwidth atomic magnetometery
  with continuous quantum nondemolition measurements},\ }\href
  {https://doi.org/10.1103/PhysRevLett.104.013601} {\bibfield  {journal}
  {\bibinfo  {journal} {Phys. Rev. Lett.}\ }\textbf {\bibinfo {volume}
  {\textbf{104}}},\ \bibinfo {pages} {013601} (\bibinfo {year}
  {2010})}\BibitemShut {NoStop}%
\bibitem [{\citenamefont {T{\'{o}}th}\ and\ \citenamefont
  {Mitchell}(2010)}]{toth2010singlet}%
  \BibitemOpen
  \bibfield  {author} {\bibinfo {author} {\bibfnamefont {G.}~\bibnamefont
  {T{\'{o}}th}}\ and\ \bibinfo {author} {\bibfnamefont {M.~W.}\ \bibnamefont
  {Mitchell}},\ }\bibfield  {title} {\bibinfo {title} {Generation of
  macroscopic singlet states in atomic ensembles},\ }\href
  {https://doi.org/10.1088/1367-2630/12/5/053007} {\bibfield  {journal}
  {\bibinfo  {journal} {New Journal of Physics}\ }\textbf {\bibinfo {volume}
  {\textbf{12}}},\ \bibinfo {pages} {053007} (\bibinfo {year}
  {2010})}\BibitemShut {NoStop}%
\bibitem [{\citenamefont {Behbood}\ \emph {et~al.}(2013)\citenamefont
  {Behbood}, \citenamefont {Colangelo}, \citenamefont {Martin~Ciurana},
  \citenamefont {Napolitano}, \citenamefont {Sewell},\ and\ \citenamefont
  {Mitchell}}]{BehboodPRL2013}%
  \BibitemOpen
  \bibfield  {author} {\bibinfo {author} {\bibfnamefont {N.}~\bibnamefont
  {Behbood}}, \bibinfo {author} {\bibfnamefont {G.}~\bibnamefont {Colangelo}},
  \bibinfo {author} {\bibfnamefont {F.}~\bibnamefont {Martin~Ciurana}},
  \bibinfo {author} {\bibfnamefont {M.}~\bibnamefont {Napolitano}}, \bibinfo
  {author} {\bibfnamefont {R.~J.}\ \bibnamefont {Sewell}},\ and\ \bibinfo
  {author} {\bibfnamefont {M.~W.}\ \bibnamefont {Mitchell}},\ }\bibfield
  {title} {\bibinfo {title} {Feedback cooling of an atomic spin ensemble},\
  }\href {https://doi.org/10.1103/PhysRevLett.111.103601} {\bibfield  {journal}
  {\bibinfo  {journal} {Physical Review Letters}\ }\textbf {\bibinfo {volume}
  {111}},\ \bibinfo {pages} {103601} (\bibinfo {year} {2013})}\BibitemShut
  {NoStop}%
\bibitem [{\citenamefont {Behbood}\ \emph {et~al.}(2014)\citenamefont
  {Behbood}, \citenamefont {Martin~Ciurana}, \citenamefont {Colangelo},
  \citenamefont {Napolitano}, \citenamefont {T\'oth}, \citenamefont {Sewell},\
  and\ \citenamefont {Mitchell}}]{BehboodPRL2014}%
  \BibitemOpen
  \bibfield  {author} {\bibinfo {author} {\bibfnamefont {N.}~\bibnamefont
  {Behbood}}, \bibinfo {author} {\bibfnamefont {F.}~\bibnamefont
  {Martin~Ciurana}}, \bibinfo {author} {\bibfnamefont {G.}~\bibnamefont
  {Colangelo}}, \bibinfo {author} {\bibfnamefont {M.}~\bibnamefont
  {Napolitano}}, \bibinfo {author} {\bibfnamefont {G.}~\bibnamefont {T\'oth}},
  \bibinfo {author} {\bibfnamefont {R.~J.}\ \bibnamefont {Sewell}},\ and\
  \bibinfo {author} {\bibfnamefont {M.~W.}\ \bibnamefont {Mitchell}},\
  }\bibfield  {title} {\bibinfo {title} {Generation of macroscopic singlet
  states in a cold atomic ensemble},\ }\href
  {https://doi.org/10.1103/PhysRevLett.113.093601} {\bibfield  {journal}
  {\bibinfo  {journal} {Physical Review Letters}\ }\textbf {\bibinfo {volume}
  {113}},\ \bibinfo {pages} {093601} (\bibinfo {year} {2014})}\BibitemShut
  {NoStop}%
\bibitem [{\citenamefont {Happer}\ and\ \citenamefont
  {Mathur}(1967)}]{PhysRev.163.12}%
  \BibitemOpen
  \bibfield  {author} {\bibinfo {author} {\bibfnamefont {W.}~\bibnamefont
  {Happer}}\ and\ \bibinfo {author} {\bibfnamefont {B.~S.}\ \bibnamefont
  {Mathur}},\ }\bibfield  {title} {\bibinfo {title} {Effective operator
  formalism in optical pumping},\ }\href
  {https://doi.org/10.1103/PhysRev.163.12} {\bibfield  {journal} {\bibinfo
  {journal} {Phys. Rev.}\ }\textbf {\bibinfo {volume} {\textbf{163}}},\
  \bibinfo {pages} {12} (\bibinfo {year} {1967})}\BibitemShut {NoStop}%
\bibitem [{Note2()}]{Note2}%
  \BibitemOpen
  \bibinfo {note} {For a derivation of the rotation angle formula see for
  example, supplementary material of \cite
  {PhysRevLett.120.040503}.}\BibitemShut {Stop}%
\bibitem [{\citenamefont {Baragiola}\ \emph {et~al.}(2014)\citenamefont
  {Baragiola}, \citenamefont {Norris}, \citenamefont {Monta\~no}, \citenamefont
  {Mickelson}, \citenamefont {Jessen},\ and\ \citenamefont
  {Deutsch}}]{PhysRevA.89.033850}%
  \BibitemOpen
  \bibfield  {author} {\bibinfo {author} {\bibfnamefont {B.~Q.}\ \bibnamefont
  {Baragiola}}, \bibinfo {author} {\bibfnamefont {L.~M.}\ \bibnamefont
  {Norris}}, \bibinfo {author} {\bibfnamefont {E.}~\bibnamefont {Monta\~no}},
  \bibinfo {author} {\bibfnamefont {P.~G.}\ \bibnamefont {Mickelson}}, \bibinfo
  {author} {\bibfnamefont {P.~S.}\ \bibnamefont {Jessen}},\ and\ \bibinfo
  {author} {\bibfnamefont {I.~H.}\ \bibnamefont {Deutsch}},\ }\bibfield
  {title} {\bibinfo {title} {Three-dimensional light-matter interface for
  collective spin squeezing in atomic ensembles},\ }\href
  {https://doi.org/10.1103/PhysRevA.89.033850} {\bibfield  {journal} {\bibinfo
  {journal} {Phys. Rev. A}\ }\textbf {\bibinfo {volume} {89}},\ \bibinfo
  {pages} {033850} (\bibinfo {year} {2014})}\BibitemShut {NoStop}%
\bibitem [{\citenamefont {Hammerer}\ \emph {et~al.}(2010)\citenamefont
  {Hammerer}, \citenamefont {S\o{}rensen},\ and\ \citenamefont
  {Polzik}}]{RevModPhys.82.1041}%
  \BibitemOpen
  \bibfield  {author} {\bibinfo {author} {\bibfnamefont {K.}~\bibnamefont
  {Hammerer}}, \bibinfo {author} {\bibfnamefont {A.~S.}\ \bibnamefont
  {S\o{}rensen}},\ and\ \bibinfo {author} {\bibfnamefont {E.~S.}\ \bibnamefont
  {Polzik}},\ }\bibfield  {title} {\bibinfo {title} {Quantum interface between
  light and atomic ensembles},\ }\href
  {https://link.aps.org/doi/10.1103/RevModPhys.82.1041} {\bibfield  {journal}
  {\bibinfo  {journal} {Rev. Mod. Phys.}\ }\textbf {\bibinfo {volume}
  {\textbf{82}}} (\bibinfo {year} {2010})}\BibitemShut {NoStop}%
\bibitem [{\citenamefont {Colangelo}\ \emph {et~al.}(2017)\citenamefont
  {Colangelo}, \citenamefont {Ciurana}, \citenamefont {Bianchet}, \citenamefont
  {Sewell},\ and\ \citenamefont {Mitchell}}]{Colangelo2017}%
  \BibitemOpen
  \bibfield  {author} {\bibinfo {author} {\bibfnamefont {G.}~\bibnamefont
  {Colangelo}}, \bibinfo {author} {\bibfnamefont {F.~M.}\ \bibnamefont
  {Ciurana}}, \bibinfo {author} {\bibfnamefont {L.~C.}\ \bibnamefont
  {Bianchet}}, \bibinfo {author} {\bibfnamefont {R.~J.}\ \bibnamefont
  {Sewell}},\ and\ \bibinfo {author} {\bibfnamefont {M.~W.}\ \bibnamefont
  {Mitchell}},\ }\bibfield  {title} {\bibinfo {title} {Simultaneous tracking of
  spin angle and amplitude beyond classical limits},\ }\href
  {https://doi.org/10.1038/nature21434} {\bibfield  {journal} {\bibinfo
  {journal} {Nature}\ }\textbf {\bibinfo {volume} {543}},\ \bibinfo {pages}
  {525} (\bibinfo {year} {2017})}\BibitemShut {NoStop}%
\bibitem [{\citenamefont {Colangelo}\ \emph {et~al.}(2013)\citenamefont
  {Colangelo}, \citenamefont {Sewell}, \citenamefont {Behbood}, \citenamefont
  {Ciurana}, \citenamefont {Triginer},\ and\ \citenamefont
  {Mitchell}}]{colangelo2013quantum}%
  \BibitemOpen
  \bibfield  {author} {\bibinfo {author} {\bibfnamefont {G.}~\bibnamefont
  {Colangelo}}, \bibinfo {author} {\bibfnamefont {R.~J.}\ \bibnamefont
  {Sewell}}, \bibinfo {author} {\bibfnamefont {N.}~\bibnamefont {Behbood}},
  \bibinfo {author} {\bibfnamefont {F.~M.}\ \bibnamefont {Ciurana}}, \bibinfo
  {author} {\bibfnamefont {G.}~\bibnamefont {Triginer}},\ and\ \bibinfo
  {author} {\bibfnamefont {M.~W.}\ \bibnamefont {Mitchell}},\ }\bibfield
  {title} {\bibinfo {title} {Quantum atom{\textendash}light interfaces in the
  gaussian description for spin-1 systems},\ }\href
  {https://doi.org/10.1088/1367-2630/15/10/103007} {\bibfield  {journal}
  {\bibinfo  {journal} {New Journal of Physics}\ }\textbf {\bibinfo {volume}
  {\textbf{15}}},\ \bibinfo {pages} {103007} (\bibinfo {year}
  {2013})}\BibitemShut {NoStop}%
\bibitem [{\citenamefont {Roy}\ \emph {et~al.}(2015)\citenamefont {Roy},
  \citenamefont {Yang}, \citenamefont {Crooker},\ and\ \citenamefont
  {Sinitsyn}}]{roy2015cross}%
  \BibitemOpen
  \bibfield  {author} {\bibinfo {author} {\bibfnamefont {D.}~\bibnamefont
  {Roy}}, \bibinfo {author} {\bibfnamefont {L.}~\bibnamefont {Yang}}, \bibinfo
  {author} {\bibfnamefont {S.~A.}\ \bibnamefont {Crooker}},\ and\ \bibinfo
  {author} {\bibfnamefont {N.~A.}\ \bibnamefont {Sinitsyn}},\ }\bibfield
  {title} {\bibinfo {title} {Cross-correlation spin noise spectroscopy of
  heterogeneous interacting spin systems},\ }\bibfield  {journal} {\bibinfo
  {journal} {Scientific Reports}\ }\textbf {\bibinfo {volume} {\textbf{5}}},\
  \href {https://doi.org/10.1038/srep09573} {10.1038/srep09573} (\bibinfo
  {year} {2015})\BibitemShut {NoStop}%
\bibitem [{\citenamefont {Lucivero}\ \emph
  {et~al.}(2017{\natexlab{b}})\citenamefont {Lucivero}, \citenamefont
  {McDonough}, \citenamefont {Dural},\ and\ \citenamefont
  {Romalis}}]{PhysRevA.96.062702}%
  \BibitemOpen
  \bibfield  {author} {\bibinfo {author} {\bibfnamefont {V.~G.}\ \bibnamefont
  {Lucivero}}, \bibinfo {author} {\bibfnamefont {N.~D.}\ \bibnamefont
  {McDonough}}, \bibinfo {author} {\bibfnamefont {N.}~\bibnamefont {Dural}},\
  and\ \bibinfo {author} {\bibfnamefont {M.~V.}\ \bibnamefont {Romalis}},\
  }\bibfield  {title} {\bibinfo {title} {Correlation function of spin noise due
  to atomic diffusion},\ }\href {https://doi.org/10.1103/PhysRevA.96.062702}
  {\bibfield  {journal} {\bibinfo  {journal} {Phys. Rev. A}\ }\textbf {\bibinfo
  {volume} {96}},\ \bibinfo {pages} {062702} (\bibinfo {year}
  {2017}{\natexlab{b}})}\BibitemShut {NoStop}%
\bibitem [{Note3()}]{Note3}%
  \BibitemOpen
  \bibinfo {note} {A detailed, quantitative treatment of the polarizability
  Hamiltonian can be found in \cite {PhysRev.163.12}. Using the equations
  presented in that reference (Eqs. III.1-III.3), we found that in all of the
  scenarios studied in the manuscript, higher spin moments affect the atomic
  susceptibility, and correspondingly the optical readout, at a level of less
  than $2 \%$. \par In principle, the effects of the optical absorption close
  to the resonance can be kept to be very small, either by increasing the area
  of the beam or by reducing the photon flux, thus rendering insignificant the
  perturbation to the spin dynamics due to light-absorption}\BibitemShut
  {NoStop}%
\bibitem [{\citenamefont {Wasilewski}\ \emph {et~al.}(2010)\citenamefont
  {Wasilewski}, \citenamefont {Jensen}, \citenamefont {Krauter}, \citenamefont
  {Renema}, \citenamefont {Balabas},\ and\ \citenamefont
  {Polzik}}]{PhysRevLett.104.133601}%
  \BibitemOpen
  \bibfield  {author} {\bibinfo {author} {\bibfnamefont {W.}~\bibnamefont
  {Wasilewski}}, \bibinfo {author} {\bibfnamefont {K.}~\bibnamefont {Jensen}},
  \bibinfo {author} {\bibfnamefont {H.}~\bibnamefont {Krauter}}, \bibinfo
  {author} {\bibfnamefont {J.~J.}\ \bibnamefont {Renema}}, \bibinfo {author}
  {\bibfnamefont {M.~V.}\ \bibnamefont {Balabas}},\ and\ \bibinfo {author}
  {\bibfnamefont {E.~S.}\ \bibnamefont {Polzik}},\ }\bibfield  {title}
  {\bibinfo {title} {Quantum noise limited and entanglement-assisted
  magnetometry},\ }\href {https://doi.org/10.1103/PhysRevLett.104.133601}
  {\bibfield  {journal} {\bibinfo  {journal} {Phys. Rev. Lett.}\ }\textbf
  {\bibinfo {volume} {104}},\ \bibinfo {pages} {133601} (\bibinfo {year}
  {2010})}\BibitemShut {NoStop}%
\bibitem [{\citenamefont {Dellis}\ \emph {et~al.}(2014)\citenamefont {Dellis},
  \citenamefont {Loulakis},\ and\ \citenamefont {Kominis}}]{Dellis}%
  \BibitemOpen
  \bibfield  {author} {\bibinfo {author} {\bibfnamefont {A.~T.}\ \bibnamefont
  {Dellis}}, \bibinfo {author} {\bibfnamefont {M.}~\bibnamefont {Loulakis}},\
  and\ \bibinfo {author} {\bibfnamefont {I.~K.}\ \bibnamefont {Kominis}},\
  }\bibfield  {title} {\bibinfo {title} {Spin-noise correlations and spin-noise
  exchange driven by low-field spin-exchange collisions},\ }\href
  {https://link.aps.org/doi/10.1103/PhysRevA.90.032705} {\bibfield  {journal}
  {\bibinfo  {journal} {Phys. Rev. A}\ }\textbf {\bibinfo {volume}
  {\textbf{90}}},\ \bibinfo {pages} {032705} (\bibinfo {year}
  {2014})}\BibitemShut {NoStop}%
\bibitem [{\citenamefont {Lucivero}\ \emph {et~al.}(2016)\citenamefont
  {Lucivero}, \citenamefont {Jim\'enez-Mart\'{\i}nez}, \citenamefont {Kong},\
  and\ \citenamefont {Mitchell}}]{Squeezed-Spin-Noise}%
  \BibitemOpen
  \bibfield  {author} {\bibinfo {author} {\bibfnamefont {V.~G.}\ \bibnamefont
  {Lucivero}}, \bibinfo {author} {\bibfnamefont {R.}~\bibnamefont
  {Jim\'enez-Mart\'{\i}nez}}, \bibinfo {author} {\bibfnamefont
  {J.}~\bibnamefont {Kong}},\ and\ \bibinfo {author} {\bibfnamefont {M.~W.}\
  \bibnamefont {Mitchell}},\ }\bibfield  {title} {\bibinfo {title}
  {Squeezed-light spin noise spectroscopy},\ }\href
  {https://link.aps.org/doi/10.1103/PhysRevA.93.053802} {\bibfield  {journal}
  {\bibinfo  {journal} {Phys. Rev. A}\ }\textbf {\bibinfo {volume}
  {\textbf{93}}},\ \bibinfo {pages} {053802} (\bibinfo {year}
  {2016})}\BibitemShut {NoStop}%
\bibitem [{\citenamefont {Vasilakis}\ \emph {et~al.}(2011)\citenamefont
  {Vasilakis}, \citenamefont {Shah},\ and\ \citenamefont
  {Romalis}}]{PhysRevLett.106.143601}%
  \BibitemOpen
  \bibfield  {author} {\bibinfo {author} {\bibfnamefont {G.}~\bibnamefont
  {Vasilakis}}, \bibinfo {author} {\bibfnamefont {V.}~\bibnamefont {Shah}},\
  and\ \bibinfo {author} {\bibfnamefont {M.~V.}\ \bibnamefont {Romalis}},\
  }\bibfield  {title} {\bibinfo {title} {Stroboscopic backaction evasion in a
  dense alkali-metal vapor},\ }\href
  {https://doi.org/10.1103/PhysRevLett.106.143601} {\bibfield  {journal}
  {\bibinfo  {journal} {Phys. Rev. Lett.}\ }\textbf {\bibinfo {volume} {106}},\
  \bibinfo {pages} {143601} (\bibinfo {year} {2011})}\BibitemShut {NoStop}%
\bibitem [{\citenamefont {Han}\ and\ \citenamefont
  {Kloeden}(2017)}]{HanBook2017}%
  \BibitemOpen
  \bibfield  {author} {\bibinfo {author} {\bibfnamefont {X.}~\bibnamefont
  {Han}}\ and\ \bibinfo {author} {\bibfnamefont {P.}~\bibnamefont {Kloeden}},\
  }\bibfield  {title} {\bibinfo {title} {Random ordinary differential equations
  and their numerical solution},\ }\href
  {https://books.google.es/books?id=Zrg7DwAAQBAJ} {\bibfield  {journal}
  {\bibinfo  {journal} {Springer Singapore}\ }\bibinfo {series} {Probability
  Theory and Stochastic Modelling} (\bibinfo {year} {2017})}\BibitemShut
  {NoStop}%
\end{thebibliography}%

\end{document}